\documentclass[fleqn]{mnras}

\usepackage{newtxtext,newtxmath}

\usepackage[T1]{fontenc}
\usepackage{lineno}

\DeclareRobustCommand{\VAN}[3]{#2}
\let\VANthebibliography\thebibliography
\def\thebibliography{\DeclareRobustCommand{\VAN}[3]{##3}\VANthebibliography}

\usepackage{graphicx}	
\usepackage{amsmath}	
\usepackage{array}
\usepackage{longtable}

\newcommand\T {{\widetilde {T}}}

\title[Physical properties of X-ray bright thermal TDEs]{From X-rays to physical parameters: a comprehensive analysis of thermal tidal disruption event X-ray spectra}

 \author[Mummery et. al.]{Andrew Mummery$^{1}$, Thomas Wevers${}^2$, Richard Saxton${}^3$ and Dheeraj Pasham${}^4$   
\\
$^{1}$ Oxford Theoretical Physics, Beecroft Building,  Clarendon Laboratory, Parks Road, Oxford, OX1 3PU, United Kingdom \\
${}^2$ European Southern Observatory, Alonso de C{\'o}rdova 3107, Vitacura, Santiago, Chile \\
${}^3$ Telespazio UK for ESA, European Space Astronomy Centre, Operations Department, 28691 Villanueva de la Cañada, Spain \\
${}^4$ Kavli Institute for Astrophysics and Space Research, Massachusetts Institute of Technology, Cambridge, MA, USA \\
}

\date{Accepted XXX. Received YYY; in original form ZZZ}

\pubyear{2022}

\begin{document}
\label{firstpage}
\pagerange{\pageref{firstpage}--\pageref{lastpage}}
\maketitle

\begin{abstract}
We perform a comprehensive analysis of a population of 19 X-ray bright tidal disruption events (TDEs), fitting their  X-ray spectra with a new, physically self consistent,  relativistic accretion disc model. Not all of the TDEs inhabit regions of parameter space where the model is valid, or have sufficient data for a detailed analysis, and physically interpretable parameters for a sub-sample of 11 TDEs are determined. These sources have thermal (power-law free) X-ray spectra.  The radial sizes measured from these spectra lie at values consistent with the inner-most stable circular orbit  of black holes with masses given by the $M_{\rm BH}-\sigma$ relationship, and  can be used as an independent measurement of $M_{\rm BH}$. The bolometric disc luminosity can also be inferred from X-ray data. 
All of the TDEs have luminosities which are sub-Eddington ($L_{\rm bol, disc} \lesssim L_{\rm edd}$), and larger than the typical hard-state transitional luminosity of X-ray binary discs ($L_{\rm bol, disc} \gtrsim 0.01 L_{\rm edd}$). The {\it peak} bolometric luminosity is found to be linearly correlated with the $M_{\rm BH}-\sigma$ mass. The TDE X-ray-to-bolometric correction can reach values up to $\sim 100$, and grows exponentially at late times, resolving the missing energy problem. We show that the peak disc luminosities of some TDEs are smaller than their observed optical  luminosities, 
implying that not all of the early time optical emission can be sourced from reprocessed disc emission. 
 Our results are supportive of the hypothesis that thermal X-ray bright TDEs are in accretion states analogous to the ``soft'' accretion state of X-ray binaries, and that black hole accretion processes are scale (mass) invariant.    
\end{abstract}




\section{Introduction}
The tidal disruption, and subsequent accretion, of a star by the super massive black hole (SMBH) at the centre of a galaxy offers a novel probe of an otherwise quiescent population of massive black holes. These tidal disruption events (TDEs) result in luminous flares which in the last two decades have been observed across a wide range of observing frequencies, including hard X-rays (e.g. Cenko et al. 2012), soft X-rays (e.g. Greiner et al. 2000), optical and UV (e.g. Gezari et al. 2008, van Velzen et al. 2021), infrared (e.g. Jiang et al. 2016, van Velzen et al. 2016b), and radio (e.g. Alexander et al. 2016, Goodwin et al. 2022). TDEs harbouring powerful radio and X-ray bright jets have also been discovered (e.g. Burrows et al. 2011). The emission from these sources typically rises to its peak in a matter of weeks to months, before fading away over the subsequent months and years.  

In recent years the number of TDEs observed at soft X-ray energies has greatly increased, with a current population of approximately 20 sources (Saxton {\it et al}. 2021). X-ray bright TDEs represent a particularly interesting TDE sub-population, as physical parameters of a TDE system can in principal be inferred from the modelling of their X-ray spectral energy distribution (Mummery \& Balbus 2020, Wen {\it et al}. 2020). The spectral energy distributions of X-ray observations of TDEs have historically been modelled with a single temperature blackbody function (e.g., Brown et al. 2017, Holoien
et al. 2018, van Velzen et al. 2019, Wevers et al. 2019, Stein et al.
2020, Cannizzaro et al. 2021, Hinkle et al. 2021; although see Wen {\it et al}. 2020 for a more realistic slim disc model), which allows two parameters to be inferred: the temperature and `size' of an emitting region.  The inferred size of the X-ray emitting region is a parameter of interest as, assuming that this emission results from a disc with inner edge at the innermost stable circular orbit (ISCO), it can be used as an estimate for the TDEs central black hole mass, $R_{\rm model} \simeq R_{\rm ISCO} \propto GM_{\rm BH}/c^2$. The black hole mass at the centre of a TDE is an important physical parameter, as it strongly correlates with the peak luminosity of the disc which forms in the aftermath of a TDE (Mummery \& Balbus 2021).  Therefore, the radial size of an X-ray bright TDE disc could, if measured correctly, be used to understand more general properties of both the TDE and the SMBH.  

In a recent paper Mummery (2021) demonstrated that the parameters inferred from fitting a single temperature blackbody model to a spectrum produced by a multi-temperature accretion disc would, for the typical parameter space of interest for TDEs, suffer from substantial systematic errors. This prevents these parameters from being interpreted physically (e.g., Wevers {\it et al}. 2019, Gezari 2021),  and inferences cannot therefore be made about the more general properties of these sources from their X-ray spectra.   Mummery (2021)  put forward an alternative X-ray spectral model, derived in the context of  relativistic thin discs, which does not suffer from these systematic errors.  This new model takes as input two parameters of physical interest, a radial scale $R_p$, and a temperature scale $T_p$. 

As the parameters inferred from fitting this new model to X-ray spectral data correspond to physical properties of the TDE disc system, self consistent inferences about the properties of X-ray emitting TDE discs can be made for the first time.  Of particular relevance is the possibility of computing the bolometric luminosities of  TDE accretion discs directly, and self consistently, from their X-ray spectrum. This  luminosity will be closely related to the disc's accretion state, which determines the fundamental physical and observable properties of the flow. 

It is the broader goal of this work to test the hypothesis that the thermal X-ray emission observed from many TDE candidates stems from accretion discs in ``soft'' accretion states similar to those observed in Galactic X-ray binary systems. This will be done by both testing whether the inferred model parameters make physical sense (i.e., do the radial scales inferred track the expected ISCO location),  and by inferring the Eddington ratios of this TDE population which, if the hypothesis is correct, should lie in the range $0.01 \lesssim f_{\rm edd} \lesssim 1.0$. 

This hypothesis is premised on the supposed scale (i.e., black hole mass) invariance of the black hole accretion process. TDEs represent an excellent probe of the scale-invariance of black hole accretion, as they evolve on much shorter timescales than  active galactic nuclei. Evidence in favor, or against, the scale invariance of black hole accretion would be a result of fundamental theoretical interest.  In addition, the development of a mapping between TDE physics and the better studied properties of X-ray binaries would allow for  a deeper understanding of the dominant emission processes of TDEs. 

To test this hypothesis, in this paper we fit the new Mummery (2021) X-ray model to the X-ray spectra of  a comprehensive sample of X-ray bright  TDEs. We find strong evidence that thermal TDEs behave as ``scaled up'' analogues of Galactic X-ray binaries, and that at least some black hole accretion states are mass independent. The radial sizes measured from TDE X-ray spectra lie at those scales expected from the black hole mass inferred from the TDE's $M-\sigma$ relationship, in contrast with previous studies using pure blackbody models. Quantitatively, the amplitudes of these radial sizes typically correspond to $1-10$ times the gravitational radius of the $M-\sigma$ black hole mass.  In addition, using both the temperature and radial scale of the inner regions of the TDE discs, we show that the bolometric luminosity of these sources peak at levels which are sub-Eddington, but larger than the typical hard state transitional scales observed in Galactic X-ray binaries. At late times we show that the bolometric luminosity of these TDEs remains exponentially larger than their observed X-ray luminosity. This provides a natural solution of the missing energy problem.   

Finally, we show that the peak disc luminosities of some of our sample are smaller than their optical peak luminosities, implying that not all of the early time optical and UV emission of all TDEs can be sourced from reprocessed disc emission, and an additional energy source is required. This could be in the form of shocks in the disc circularisation process (as argued by e.g., Shiokawa et al. 2015, Piran et al.
2015, Bonnerot \& Stone 2021, Bonnerot et al. 2021). 

The layout of this paper is as follows. In section \ref{sec2},   we recap the disc model introduced in Mummery (2021). In section \ref{sec3} we present the TDE sample used in this study; we discuss the fitting techniques and procedure in section \ref{sec4}.  In section \ref{sec5} we analyse the results of the spectral fits, before concluding in section \ref{conc}.  Some additional data tables and figures are presented in Appendices. Optical spectroscopic redshifts are available for all of the sources in our sample, and these are converted to luminosity distances assuming a flat $\Lambda$ cold dark matter cosmology with $H_0$ = 70 km s$^{-1}$ Mpc$^{-1}$, $\Omega_m$ = 0.3 and $\Omega_\Lambda = 0.7$.

\section{The new model}\label{sec2}
An accretion disc in the ``soft'' state produces broadband emission well described by a superposition of black bodies. Each disc region has both a different emitting temperature and different emitting area, and contributes its own blackbody spectrum to the gross disc spectrum. In a relativistic disc system the observed emission is modified by both gravitational red-shifting, and by the Doppler boosting of the local emission by the disc fluids relativistic rotational velocities. Finally, the opacity of the disc atmosphere may differ from that of pure absorption, with electron scattering important in many regimes. When scattering of photons is dominant, the local emission will appear hotter than expected {(Shimura and Takahara 1995, Done et al. 2012)}, as photons which originated deeper in the disc are observed {(Davis et al. 2006)}. 

Combining these effects, the observed thermal  emission from an accretion disc is generally given by the following expression 
\begin{equation}\label{fullmodel}
F_\nu =  \frac{1}{D^2}\iint_{\cal S} {f_\gamma^3 f_{\rm col}^{-4} B_\nu (\nu/f_\gamma, f_{\rm col} T)} ~\text{d}b_1\text{d} b_2.
\end{equation}
Here ${\cal S}$ is the surface of the disc, and $b_1$ and $b_2$ are cartesian image plane photon impact parameters. The function $B_\nu$ is the Planck function:
\begin{equation}
B_\nu(\nu, T)  =  \frac{2\pi h \nu^3}{ c^2} \frac{1 }{ \exp\left({h\nu / kT}\right) - 1} .
\end{equation}
The factor $f_\gamma$ is the photon energy-shift factor, defined as the ratio of the observed photon frequency $\nu$ to the emitted photon frequency in the discs rest frame $\nu_{\rm emit}$, $f_\gamma \equiv \nu/\nu_{\rm emit}$.  The constants $h, k$ and $c$ are the Planck constant, Boltzmann constant, and the speed of light respectively.  $T$ is the temperature of the disc, a function of both disc radius and time $T(r,t)$. Finally, $f_{\rm col}$ is the `color-correction' factor, which is included to model disc {opacity} effects. This correction factor generally takes a value $f_{\rm col} \sim 2.3$ for typical TDE disc temperatures (Done {\it et al}. 2012). The parameter $D$ is the luminosity distance to the source.  For a more detailed derivation and discussion of this expression we refer to Section 2 of Mummery \& Balbus 2021. 

While this expression may look unwieldy, many TDE discs have the fortunate property of being relatively cool, with their spectra peaking below the lower bandpass of X-ray telescopes, $kT  \ll 0.3$ keV. This means that X-ray observations of TDE  discs probe the quasi-Wien tail of the disc spectrum, a limit in which equation (\ref{fullmodel}) becomes analytically tractable. In Mummery \& Balbus (2021) it was shown that equation (\ref{fullmodel}) can be integrated by performing a Laplace expansion about the hottest regions of the disc.  

A Laplace expansion proceeds by defining a small parameter, in our case 
\begin{equation}
\varepsilon \equiv {k\T_p \over h \nu} \ll 1,
\end{equation}
where $T_p$ is the peak temperature of the disc at a given time.  For $\varepsilon \ll 1$, the Planck function can be replaced by its Wien-tail form 
\begin{equation}
    B_\nu(\nu, T)  \simeq  \frac{2\pi h \nu^3}{ c^2}  \exp\left(-{h\nu \over kT}\right)  ,
\end{equation}
an approximation which introduces exponentially small corrections. 

In the limit $\varepsilon \ll 1$, the Wien-tail flux observed across an X-ray band pass results physically from only a very small region of the disc, with the contributions from all other disc regions exponentially suppressed. We may therefore spatially Taylor expand the temperature profile in the exponential, and perform the integral (equation \ref{fullmodel}) term by term. The resulting Laplace expansion has the form (Mummery \& Balbus 2021)
\begin{equation}\label{LaplaceExpansion}
    F_\nu(\varepsilon)  = F_0 \,\varepsilon^\gamma\, \exp\left(-{1 \over \varepsilon}\right) \,  \sum_{n=0}^{\infty} \, \alpha_n \varepsilon^n ,
\end{equation}
where $F_0$ and $\alpha_n$ are temperature-independent constants, and $\gamma$ depends on the precise properties of the temperature maximum (this will be further discussed below). 

A key mathematical point here is that the X-ray spectral model built upon this Laplace expansion has only a finite parameter space of applicability, and can only be used when $kT_p \ll h\nu$. The corrections involved in approximating the solution of equation (\ref{fullmodel}) with equation (\ref{LaplaceExpansion}) are exponentially small in the small $\varepsilon$ limit. However, these corrections are not small in the opposite limit, and this model will quickly breakdown when $\varepsilon \sim 1$. The corrections to this model in the $\varepsilon > 1$ limit cannot  be mitigated by keeping additional terms in the summation, which is not guaranteed to converge for $\varepsilon > 1$. On one hand, if one fits an X-ray model derived in this manner to data and finds a set of best fit parameters which imply $\varepsilon \sim 1$ then, even if the fit is formally acceptable, the inferred parameters cannot be physically interpreted. On the other hand, this model will converge exponentially to the full solution for $\varepsilon < 1$ and therefore, in its regime of applicability, will be highly accurate. 

In Mummery \& Balbus (2021) the Laplace expansion solution of equation (\ref{fullmodel}) was found,  and it is given by the following expression 
\begin{multline}\label{MB}
F_\nu(R_p, \T_p)  = \frac{4\pi \xi_1 h\nu^3}{c^2f_{\rm col}^4}\left( \frac{R_p}{D} \right)^2 \left(\frac{k \T_p}{h \nu} \right)^\gamma \exp\left(- \frac{h\nu}{k \T_p} \right) \\
\times \left[ 1 + \xi_2\left(\frac{k \T_p}{h \nu} \right) + \xi_3\left(\frac{k \T_p}{h \nu} \right)^{2}   +\, \dots \right],
\end{multline} 
Here we have defined $\T_p \equiv \max(f_{\rm col} f_\gamma T)$, the maximum ``effective'' temperature in the accretion disc. The radius $R_p$ corresponds to the image plane co-ordinate of this hottest region. The constant $\gamma$ depends on assumptions about both the inclination angle of the disc and the  disc inner boundary condition, and is limited to the range $1/2 \leq \gamma \leq  3/2$. We note that $\gamma = 1/2$ for a vanishing ISCO stress disc observed precisely face on.  The positive constants $\xi_1, \xi_2$ and $\xi_3$ are all order unity, $\xi_1 \simeq 2.19, \xi_2 \simeq 3.50, \xi_3 \simeq 1.50$ (Mummery \& Balbus 2021). Here `$\dots$' denotes higher order terms which scale $\propto ({k \T_p}/{h \nu} )^{n }$ (where $n \geq 3$), which can be safely neglected, provided we are in the parameter space of interest. 

While this equation was derived in the context of relativistic thin discs, it is actually more broadly applicable than in just the ``thin'' disc limit.  The only assumptions inherent to this modelling is that each disc radius emits like a colour-corrected blackbody, and that there exists some disc radius $R_p$ where the disc temperature peaks, at a level below the observed frequency $k\T_p \ll h\nu_l$. These assumptions will still be valid in the ``slim'' disc limit, and thus this model will produce valid descriptions of the X-ray spectra of accreting sources at high Eddington ratios $f_{\rm edd} \sim 1$.

The X-ray spectrum of an accretion disc observed in the Wien-tail can therefore be described by just three parameters: $R_p$, $T_p$ and $\gamma$. Two of these parameters, $R_p$ and $T_p$, encode  physical information about the disc system itself. The value of $R_p$ will be driven primarily by the central black holes mass, for example, a parameter of real physical interest. In addition,  the bolometric luminosity of the  accretion disc will scale as $R_p^2 T_p^4$, and can therefore be constrained directly from the detailed modeling of X-ray spectra (section \ref{sec5}). 

\section{The data set}\label{sec3}
We search the literature for all promising TDE candidates for which good quality X-ray data (in the Swift/XRT, XMM-Newton and NICER archives) are publicly available. We include only those TDEs detected prior to April 2020. This yields a list of $\sim$20 sources (Table \ref{table:sample}), excluding ROSAT TDE candidates for which only ROSAT data is available (because these spectra are not retrievable).
We furthermore collect the peak integrated UV (blackbody) luminosities from the literature, as well as velocity dispersion measurements that can be converted to black hole mass estimates through the M--$\sigma$ relation. Galactic hydrogen column densities ($n_H$) are taken from the HI4PI survey (HI4PI Collaboration, 2016). 

\subsection{XMM-Newton data reduction}
Observations were downloaded from the XMM-Newton Science Archive and processed with the XMM-Newton
Science Analysis System  (SAS v19.1.0; Gabriel et al. 2004). EPIC-pn event files were filtered for background solar flares, following the standard processing thread with a cut at a count rate of 0.4 c/s in the 10-12 keV band. Source spectra were 
extracted using a circular region about the source and local background spectra were produced 
from a source-free region on the same CCD.

\subsection{NICER}
AT2019dsg and AT2020ksf were observed by {\it NICER} on multiple epochs (see Table \ref{table:results}). We started {\it NICER} data analysis with the raw level-1 files that we downloaded from the publicly accessible High Energy Astrophysics Science Archive Research Center (HEASARC; \href{https://heasarc.gsfc.nasa.gov/cgi-bin/W3Browse/w3browse.pl}{https://heasarc.gsfc.nasa.gov/cgi-bin/W3Browse/w3browse.pl}). We used the {\it nicerl2} task to extract the cleaned eventlists. To produce the Good Time Intervals (GTIs), we used the default values for all parameters except for the overshoot, the undershoot, and overshoot expression. These were screened at a later stage to minimize the loss of data. For a detailed discussion on this see (Pasham et al. 2021). X-ray spectra were filtered for so-called hot detectors and the ancillary response files (arf) and the redistribution matrix files (rmfs) were generated using the tools {\it nicerarf} and {\it nicerrmf}, respectively. The background spectra were estimated using the 3c50 model (Remillard et al. 2022).

\subsection{Swift/XRT}
We use the Swift online XRT tool to extract (stacked) X-ray spectra of all sources with sufficient coverage. If significant variability was reported (or observed) we make stacked spectra for different time ranges as indicated in Table \ref{table:1}. 

Swift X-ray spectra typically contain far lower photon counts than XMM or NICER spectra. As such we typically rely  on XMM and NICER spectra for our analysis, using Swift spectra only for those sources lacking observations with these instruments (see Table \ref{table:results}). 

\subsection{Magellan/MagE optical spectroscopy}
The host galaxy of 2XMM J1847 was observed on 19 August 2021 with the Magellan Echellete spectrograph (MagE), mounted on the Magellan Baade telescope located at Las Campanas Observatory, Chile. We used a 0.7 arcsec slit, which delivers a FWHM spectral resolution of 50 km s$^{-1}$ at 4000 \AA. The spectrum was reduced using the dedicated MagE data reduction pipeline (Kelson et al. 2000, Kelson  2003).  
Subsequently the spectrum is normalised to the continuum by fitting a low order spline function to remove the strong curvature from each échelle order, masking strong emission and absorption lines that may distort the continuum identification. 
We then follow the method of (Wevers et al. 2017)  to measure the stellar velocity dispersion by combining the penalized pixel fitting routine (Cappellari  2017) with the Elodie stellar spectral library (Prugniel et al. 2007). We measure a velocity dispersion of $\sigma = 91 \pm 4 $ km s$^{-1}$. {This can be turned into an estimate for the 2XMM J1847 black hole mass using the $M-\sigma$ relationship. Explicitly this $\sigma$ measurement implies a black hole mass of $\log_{10}M_{\rm BH}/M_\odot$ = 6.6 $\pm$ 0.4 (Ferrarese \& Ford 2005), or $\log_{10}M_{\rm BH}/M_\odot$ = 6.4 $\pm$ 0.4 (McConnell \& Ma 2013).}

\begin{table*}
    \caption{Sample used in this work together with some basic properties. Galactic column densities are taken from the HI4PI survey (HI4PI collaboration, 2016). The references are: [1] Holoien et al. 2016a, [2] Wen et al. 2020, [3] Holoien et al. 2016b, [4] Gezari et al. 2017, [5] Wevers et al. 2019a, [6] Wyrzykowski et al. 2016, [7] Kajava et al. 2020, [8] Cannizzaro et al. 2021, [9] van Velzen et al. 2019, [10] Wevers et al. 2019b, [11] Wevers 2020, [12]  Lin et al. 2017, [13] Saxton et al. 2017, [14] Lin et al. 2015, [15] Wevers et al. 2022, [16] Saxton et al. 2012, [17] Saxton et al. 2019, [18] Lin et al. 2011, [19] Short et al. 2020, [20] Wevers et al. 2019a, [21] Liu et al. 2019, [22] Nicholl et al. 2020. [23] Pasham et al. 2020, [24] Stein et al. 2021.  }

    \centering
    \begin{tabular}{cccccccc}
Source & RA & Dec. & Redshift & Distance & Galactic $n_H$ & $\sigma$  & Reference  \\
 &  &  &  & (Mpc) & (10$^{20}$ cm$^{-2}$) & (km/s) & \\\hline
 ASASSN--14li &12 48 15.23& 17 46 26.4 & 0.0206 & 90 & 1.9 & 78 $\pm$ 2 & [1], [2] \\ 
 ASASSN--15oi &20 39 09.03&--30 45 20.8& 0.051 & 216 & 4.8 & 61$\pm$7 & [3], [4], [5]\\
 OGLE16aaa &01 07 20.88 &--64 16 20.7& 0.1655 & 800 & 2.7 & --- & [6], [7] \\ 
 AT2019dsg &20 57 02.974 &+14 12 15.86& 0.0512 & 224 & 6.5 & 94 $\pm$ 1 & [8], [24]\\
 AT2018zr &07 56 54.54& +34 15 43.61& 0.071 & 322 & 4.4 & --- & [9]\\
 AT2018fyk &22 50 16.09&--44 51 53.50& 0.059 & 264 & 1.15 & 158$\pm$1 & [10], [11] \\ 
 3XMM J1500 &15 00 52.07 & +01 54 53.8 &0.145 & 692 & 4.1 & 59$\pm$3 & [12] \\
 XMMSL1 J0740 &07 40 08.09 & --85 39 31.3& 0.0173 & 73 & 5.3 & 112$\pm$ 3& [13], [11] \\ 
 3XMM J1521 &15 21 30.72 &+07 49 16.5&0.179 & 866 & 2.7 & 58$\pm$2 & [14] \\ 
 GSN069 &01 19 08.663 & --34 11 30.52& 0.018 & 69 & 2.3 &64 $\pm$ 4 & [15]  \\ 
 SDSSJ1201 &12 01 36.03 &+30 03 05.5& 0.146 & 700 & 1.3 & 122$\pm$4 &[16] \\ 
 XMMSL2 J1446 &14 46 05.22 &+68 57 31.1& 0.029 & 127 & 1.7 &167 $\pm$15& [17], [11] \\ 
 2XMM J1847 &18 47 25.12&--63 17 25.3& 0.035 & 156 & 6.3 &  91$\pm$4 & [18] \\
 AT2018hyz &10 06 50.871 &+01 41 34.08& 0.0457 & 204 & 2.7 & 57 $\pm$ 1& [19] \\ 
 RBS1032 &11 47 26.69 &+49 42 57.7& 0.026 & 114 & 1.4 & 49 $\pm$ 7 & [20] \\ 
 AT2019azh &08 13 16.95 &+22 38 54.03& 0.022 & 96 & 4.2 & 77 $\pm$ 2 & [21]\\
 2MASX J0249 &02 49 17.31 &--04 12 52.1 & 0.019 & 83 & 3.2 & 43 $\pm$ 4& [20] \\ 
 AT2019qiz &04 46 37.88 & --10 13 34.90& 0.0151 & 66 & 6.6 & 71 $\pm$ 2 & [22] \\ 
 AT2020ksf &21 35 27.26&--18 16 35.54& 0.0923 & 426 & 3.6 & --- & [23] \\
 \hline
    \end{tabular}
    \label{table:sample}
\end{table*}

\begin{table*}
    \caption{Best fit parameters obtained through spectral fitting.  Cash statistics are used to find the best fit parameters, and are indicated in the Fit stat column together with the number of degrees of freedom (d.o.f.). The uncertainties indicate the 90 per cent confidence intervals, obtained by varying the parameters and finding $\Delta$stat = 2.7. The spectral range over which the fit was performed is also listed. $^\dagger$ indicates a column density fixed to the Galactic value (see text). We only display the best fitting value of $\gamma$ in this table. The $\gamma$ parameter is poorly constrained by our fits and the uncertainty interval on each measurement should be understood to include the entire allowed range $\gamma \in 1/2$ -- $3/2$.}
    \label{table:results}
    \centering
    \renewcommand{\arraystretch}{1.2}
    \begin{tabular}{cccccccccc}
       Source  & Instrument & Photon counts & MJD & Spectral range & $n_H$ & $R_p$ & $T_p$ & $\gamma$ & Fit statistic \\   &&& (days) & (keV) & (10$^{20}$ cm$^{-2}$) & ($10^{12}$ cm) & ($10^{5}$ K) & & (stat$\big/$d.o.f.) \\\hline
    ASASSN--14li & XMM/RGS & 9200 & 56997& 0.35 -- 1.2 & $6.1^{+2.2}_{-1.5}$ & $6.1^{+3.7}_{-0.5}$ & $3.7^{+0.6}_{-0.1}$ & 0.5 & 353/266 \\
    ASASSN--14li & XMM/RGS & 36150 & 56999& 0.35 -- 1.2 & $4.6^{+0.8}_{-0.7}$ & $4.8^{+0.8}_{-0.2}$ & $3.8^{+0.2}_{-0.1}$ & 0.5 & 1267/873 \\
    ASASSN--14li & XMM/RGS & 9600 & 57024& 0.35 -- 1.2 & $6.5^{+1.4}_{-1.2}$ & $6.4^{+1.7}_{-0.4}$ & $3.6^{+0.2}_{-0.04}$ & 0.5 & 378/269 \\
    ASASSN--14li & XMM/RGS & 2086 & 57213& 0.35 -- 1.2 & $4.0^{+0.5}_{-0.5}$ & $4.8^{+4.5}_{-0.7}$ & $3.2^{+0.6}_{-0.1}$ & 0.5 & 79/72 \\
    ASASSN--14li & XMM/RGS & 5700 & 57367& 0.35 -- 1.2 & $4.0^{+0.5}_{-0.5}$ & $5.2^{+4.1}_{-0.7}$ & $3.0^{+0.4}_{-0.1}$ & 0.5 & 222/232 \\
    ASASSN--14li & XMM/PN & 22400 & 57399 & 0.2 -- 1.0 & $3.7^{+0.9}_{-0.4}$ & $5.1^{+1.5}_{-0.3}$ & $2.9^{+0.4}_{-0.1}$ & 0.5 & 214/157 \\
    ASASSN--14li & XMM/PN & 18400  & 57544 & 0.2 -- 1.0 & $3.9^{+0.6}_{-0.5}$ & $6.0^{+1.6}_{-0.4}$ & $2.6^{+0.2}_{-0.1}$ & 0.5 & 171/157\\
    ASASSN--14li & XMM/PN & 9800 & 57727 & 0.2 -- 1.0 & $3.9^{+0.8}_{-0.6}$ & $6.2^{+3.2}_{-0.5}$ & $2.5^{+0.3}_{-0.2}$ & 0.5 & 133/157 \\
    ASASSN--14li & XMM/PN & 8200 & 57912& 0.2 -- 1.0 & $3.7^{+0.9}_{-0.8}$ & $8.0^{+6.1}_{-0.7}$ & $2.2^{+0.4}_{-0.1}$ & 0.5 & 129/157 \\
    ASASSN--14li & XMM/PN & 7600 & 58093 & 0.2 -- 1.0 &$1.5^{+0.8}_{-0.6}$ & $2.6^{+2.4}_{-0.2}$ & $2.5^{+0.4}_{-0.2}$ & 0.5 & 175/157 \\\hline
    
    ASASSN--15oi & XMM/PN & 600 & 57324 & 0.2 -- 1.0 & $4.8^\dagger$ & $0.58^{+0.22}_{-0.16}$ & $5.0^{+0.5}_{-0.7}$ & 1.5 & 139/157 \\
    ASASSN--15oi & XMM/PN & 4050 & 57482 & 0.2 -- 1.0 & $4.8^\dagger$ &  $2.8^{+0.8}_{-0.3}$ & $3.3^{+0.3}_{-0.1}$ & 0.5 & 179/157 \\\hline
    
    OGLE16aaa & XMM/PN & 260 & 57548 & 0.2 -- 1.0 & $3.3^{+3.4}_{-3.3}$ & $1.7^{+1.4}_{-0.8}$ & $4.1^{+1.9}_{-1.2}$ & 0.5 & 112/157\\
    OGLE6aaa & XMM/PN & 4500  & 57722 & 0.2 -- 1.0 & $1.3^{+0.6}_{-0.9}$ & $1.5^{+0.5}_{-0.6}$ & $6.1^{+0.4}_{-1.1}$ & 0.5 & 173/157\\ \hline
    
    AT2019dsg & NICER & 13000 & 58624 &0.3 -- 1.0& $8.6^{+1.4}_{-0.8}$ & $1.8^{+1.3}_{-0.3}$ & $4.9^{+0.8}_{-0.1}$ & 0.5 & 10.5/13 \\
    AT2019dsg & NICER & 3800 & 58625 &0.3 -- 1.0& $10.0^{+1.7}_{-3.1}$ & $2.8^{+1.4}_{-1.4}$ & $5.6^{+0.5}_{-0.9}$ & 1.5 & 14/12 \\
    AT2019dsg & NICER & 1300 & 58630 &0.3 -- 1.0& $8.1^{+2.4}_{-0.7}$ & $1.8^{+2.8}_{-0.5}$ & $4.8^{+1.2}_{-0.5}$ & 0.5 & 8/12 \\
    AT2019dsg & NICER & 750 & 58633 &0.3 -- 1.0& $9.1^{+2.9}_{-1.1}$ & $2.3^{+4.7}_{-0.9}$ & $4.5^{+1.3}_{-0.5}$ & 1.4 & 13/12 \\
    AT2019dsg & NICER & 660 & 58634 &0.3 -- 1.0& $6.3^{+2.7}_{-0.8}$ & $1.1^{+1.9}_{-0.5}$ & $5.0^{+1.7}_{-0.6}$ & 0.6 & 10/11\\ \hline
    
    AT2018zr & XMM/PN & 300 & 58220 & 0.2 -- 1.0 & $14.3^{+2.7}_{-3.3}$ & $0.071^{+0.040}_{-0.036}$ & $9.5^{+0.3}_{-1.1}$ & 1.5 & 198/157\\
    AT2018zr & XMM/PN & 185 & 58242 & 0.2 -- 1.0 & $17.1^{+8.9}_{-5.1}$ & $0.095^{+0.045}_{-0.040}$ & $8.7^{+0.5}_{-2.2}$ & 1.5 & 165/157 \\ \hline
    
    3XMM J1521 & XMM/PN & 3050  & 51778  &0.2 -- 1.2 & $2.1^{+1.3}_{-0.8}$ & $0.32^{+0.03}_{-0.03}$ & $8.2^{+1.7}_{-0.5}$ & 0.5 & 232/198\\\hline
    
    GSN 069 & XMM/PN & 55000 & 56996 &0.2 -- 1.0& $3.3^{+0.3}_{-0.3}$ & $0.67^{+0.06}_{-0.02}$ & $3.8^{+0.2}_{-0.05}$ & 0.5 & 256/158\\\hline

    2XMM J1847 & XMM/PN &2000 & 53985 &0.2 -- 1.2& $6.3^\dagger$ & $0.20^{+0.03}_{-0.03}$ & $7.5^{+0.4}_{-0.4}$ & 1.5 & 176/175 \\
    2XMM J1847 & XMM/PN &18000 & 54206 & 0.2 -- 1.2& $9.6^{+0.4}_{-0.8}$ & $0.55^{+0.05}_{-0.07}$ & $8.2^{+0.2}_{-0.4}$ & 1.5 & 254/198 \\\hline
    
    AT2019azh & Swift &240& 58553 - 58634 &0.3 -- 10& $5.4^{+8.0}_{-5.3}$ & $0.6^{+3.8}_{-0.5}$ & $4.3^{+1.8}_{-1.5}$ & 1.5 & 90/82 \\
    AT2019azh & Swift &2500& 58767 - 58977 &0.3 -- 10& $4.2^\dagger$ & $1.4^{+1.1}_{-0.1}$ & $3.8^{+1.0}_{-0.05}$ & 0.5 & 55/65 \\\hline
    
    2MASX J0249 & XMM/PN &1780& 53930 &0.2 -- 1.2 & $31.8^{+10.2}_{-3.8}$ & $0.23^{+0.52}_{-0.09}$ & $6.8^{+1.6}_{-0.6}$ & 0.5 & 260/219 \\\hline
    
    AT2020ksf & NICER &14450& 59187 - 59189&0.3 -- 1.5& $3.3^{+1.4}_{-0.6}$ & $0.8^{+0.3}_{-0.05}$ & $7.5^{+1.0}_{-0.1}$& 0.5 & 22/22 \\
    AT2020ksf & NICER &11230& 59191 - 59195&0.3 -- 1.4& $6.2^{+1.0}_{-0.9}$ & $1.7^{+0.5}_{-0.2}$ & $5.7^{+0.4}_{-0.2}$ & 0.5 & 27/19\\
    AT2020ksf & Swift &440 &59179 - 59205 &0.3 -- 1.5& $5.2^{+6.2}_{-3.5}$ & $1.1^{+2.1}_{-0.5}$ & $6.3^{+2.1}_{-0.7}$ & 0.5 & 45/65 \\\hline

    \end{tabular}
\end{table*}

\section{Analysis}\label{sec4}
\subsection{Model implementation }
The model described by equation (\ref{MB}) takes as input four parameters, one of which -- the source-observer distance $D$ -- is fixed at the observed distance of the TDEs host galaxy. The remaining three parameters ($R_p$, $\T_p$ and the index $\gamma$) are allowed to vary for each observation. 

The index $\gamma$, which determines the leading power-law behaviour of the disc spectrum, is the only parameter which is constrained theoretically, as it must lie in the following range: 
\begin{equation}
    1/2 \leq \gamma \leq 3/2 .
\end{equation}
Physically, the parameter $\gamma$ varies depending on two factors. First, the observed geometry of the hottest region of the accretion disc (whether the disc is observed edge-on or face-on). Second, on whether the disc temperature maximum occurs at the disc inner edge, or in the bulk of the disc body. The following values are known from theory:
\begin{itemize}
    \item $\gamma = 1/2$, disc observed face on, with the temperature maximum inside the disc body $R_{\rm in} < R_p < R_{\rm out}$. 
    \item $\gamma = 1$, disc observed edge on, with the temperature maximum inside the disc body $R_{\rm in} < R_p < R_{\rm out}$. 
    \item $\gamma = 1$, disc observed face on, with the temperature maximum at the disc interior $R_{\rm in} = R_p$. 
    \item $\gamma = 3/2$, disc observed edge on, with the temperature maximum at the disc interior $R_{\rm in} = R_p$. 
\end{itemize}
The properties of $\gamma$ are discussed in more detail in Balbus (2014) and Mummery \& Balbus (2020a). In practice, the observed accretion disc spectrum is only weakly dependent on $\gamma$, which cannot be strongly constrained from observations.  In this work we therefore treat $\gamma$ as a nuisance parameter, letting it vary between its allowed limits at each epoch. In fact, the 1-$\sigma$ uncertainties on $\gamma$ typically fill the entire allowed range of $\gamma \in 1/2$ -- $3/2$, and as such the $\gamma$ parameter merely extends the uncertainty range of the parameters $R_p$ and $T_p$.  

The quantity $\T_p$ is given by $\T_p \equiv f_{\rm col} f_\gamma T_p$, where $T_p$ is the hottest physical temperature in the accretion disc. The XSPEC model takes the physical peak temperature $T_p$ as an input. We then compute the disc color-correction factor using the Done {\it et al}. (2012) model: 
\begin{align}\label{col1}
&f_{\rm col}(T_p) =  \left(\frac{72\, {\rm keV}}{k T_p}\right)^{1/9}, \quad  T_p > 1\times10^5 {\rm K}. \\
&f_{\rm col}(T_p) =  \left(\frac{T_p}{3\times10^4 {\rm K}} \right)^{0.83}, \quad 3\times10^4 {\rm K} < T_p < 1\times10^5 {\rm K} .\label{col2} \\
&f_{\rm col}(T_p) = 1, \quad T_p < 3\times 10^4 {\rm K} . \label{col3}
\end{align}
This model is routinely used for the modelling of AGN disc spectra. It is likely that the disc conditions in TDEs will be most similar to those in AGN (though AGN discs will be more radially extended), and so this model should accurately model TDE disc colour-correction effects.

The peak temperatures of TDE discs are typically in the high temperature regime $T_p > 1\times 10^5$K. The disc color-correction factor is therefore only weakly temperature dependent $f_{\rm col} \propto T_p^{-1/9}$, with typical value $f_{\rm col} \simeq 2.3$. The final component of computing $\T_p$ is the photon energy-shift factor $f_\gamma$. This quantity depends on numerous factors which lie beyond the scope of this model, chiefly the black hole spin, disc-observer inclination angle, and the radius in gravitational units at which the temperature maximum occurs. As none of these parameters can be determined from the existing observations, we use the energy-shift value associated with the ISCO radius of a Schwarzschild black hole, observed face-on $f_\gamma = 1/\sqrt{2}$. The factor $f_\gamma$ does not vary strongly over the entire parameter space of black hole spin, disc radius and observer inclination angle, and so uncertainty in its value will not cause significant errors to propagate into the inferred disc parameters (this is verified in section \ref{sec5}). 


\subsection{Fitting procedure} 
The model described above is fit to the data (as summarised in Tables \ref{table:sample}, \ref{table:results}) using Xspec v12.11.1 in HEASoft v6.28 (Arnaud 1996). We account for Galactic extinction by including Hydrogen column densities ($n_H$) as measured by HI4PI (HI4PI collaboration, 2016) and tabulated in Table \ref{table:sample}, using a {\tt TBabs} model in Xspec. Initially this is left as a free parameter, but if the best fit value is significantly below the Galactic value, we fix it to the Galactic value instead. This may happen if the signal-to-noise ratio is low, which can be assessed by either the number of photons in the spectrum or the number of degrees of freedom, both of which are  tabulated in Tables \ref{table:1},  \ref{table:results}. {We verified that the finding of sub-galactic $n_H$ was a result of poor data quality and not a model deficiency by re-fitting these spectra with the {\tt diskbb} model. We also found sub-galactic $n_H$ for {\tt diskbb}, supporting the low signal-to-noise hypothesis. } The normalisation of the disk model is set by the distance, so the model norm parameter is fixed to 1. We use the unbinned spectra in combination with Cash statistics (Cash 1979) to find the best fit parameters, unless otherwise indicated; for the NICER data we use the optimal binning scheme of Kaastra \& Bleeker (2016).

The spectral range that is used in the fitting is also listed in Table \ref{table:results}, as it can differ depending on the instrument response, signal-to-noise ratio of the data, as well as the spectral shape. For example, a pure thermal spectrum will tend to be noise-dominated for $E \gtrsim$1-1.5 keV, while a thermal + power-law spectrum will contain signal above the background out to higher energies. In the latter case, the expectation (and observation, e.g. Wevers et al. 2019) is that the thermal component dominates at low energies $\lesssim 1$ keV while the non-thermal component dominates at higher energies. For sources with a prominent power-law spectral component present we fit the data  using a joint model ({\tt tdediscspec + powerlaw}) to describe the entire energy range with signal above the noise. 


Finally, uncertainties are assessed by using the Xspec error command. In cases where this leads to divergent results, we instead perform a manual stepwise exploration of the fit statistic surface and determine 90 per cent confidence intervals by finding the parameter values for which $\Delta$ stat = 2.7, using the steppar command.
Examples of X-ray spectra and their best-fit models are shown in Figure \ref{fig:examplespectra}.

For ASASSN14li, the presence of additional ionized absorption has been reported (Kara et al. 2018). We therefore tested the effect of including an additional absorption component (using the {\tt gabs} model) in the accretion disk model parameters (primarily $n _H$, $T_p$ and $R_p$), finding that this can lead to variations of ~25 $\%$ in these parameters. This is typically within the reported error budget.

The best fit parameters and their uncertainties are presented in Table \ref{table:results}. Some example best-fitting X-ray spectra are displayed in Figure \ref{fig:examplespectra} in Appendix \ref{secB}. 

\subsection{Sources which must be removed from the sample}
We reiterate that the corrections to equation (\ref{MB}) are large and poorly constrained when the disc temperature is of order the bandpass energy $k\T_p \sim E_l$. We  cannot mitigate this effect by keeping additional higher order turns in the Laplace expansion (equation \ref{LaplaceExpansion}), and must instead simply remove these sources from our sample.

The best fitting parameter values inferred from the spectral fits to the sources XMMSL1 J0740, XMMSL2 J1446, 3XMM J1500, AT2018fyk and SDSS J1201 all lie outside of the regime of validity of the derivation of the underlying physical model (equation \ref{MB}). Unfortunately, this means that the parameters inferred for these sources cannot be physically interpreted.

Many of the sources which have inferred temperatures which are ``too high'' to be interpreted within our model also have a strong power-law component present in their spectra, in addition to a thermal component (e.g. AT2018fyk, XMMSL1 J0740, XMMSL2 J1446). It is likely that these temperature values are not physical, and are merely an artifact of fitting a thermal model added to a distinct power law component to a system where the thermal and nonthermal components are likely to be intrinsically linked. The addition of a distinct power-law component will only ever approximate the effect of Comptonisation on the Wien spectrum, and the model temperature of these sources is likely strongly biased upwards by these simple model artifacts. 

Three sources in our sample, AT2018hyz, AT2019qiz, and RBS1032 have a very low number of photon counts: 50 (AT2018hyz), 65 (AT2019qiz) and 120 (RBS1032) respectively. For these sources it was not possible to compute error ranges on the best fitting parameters, the fit was simply too degenerate, and we remove them from our sample on data quality grounds. 

As these eight sources are either not in the regime where the observed emission is dominated by the Wien-tail of thermal disc emission, or do not have sufficient photon counts for a robust statistical analysis, we do not include them in our analysis going forward.


\section{Results and implications}\label{sec5}
\begin{figure}
    \centering
    \includegraphics[width=\linewidth]{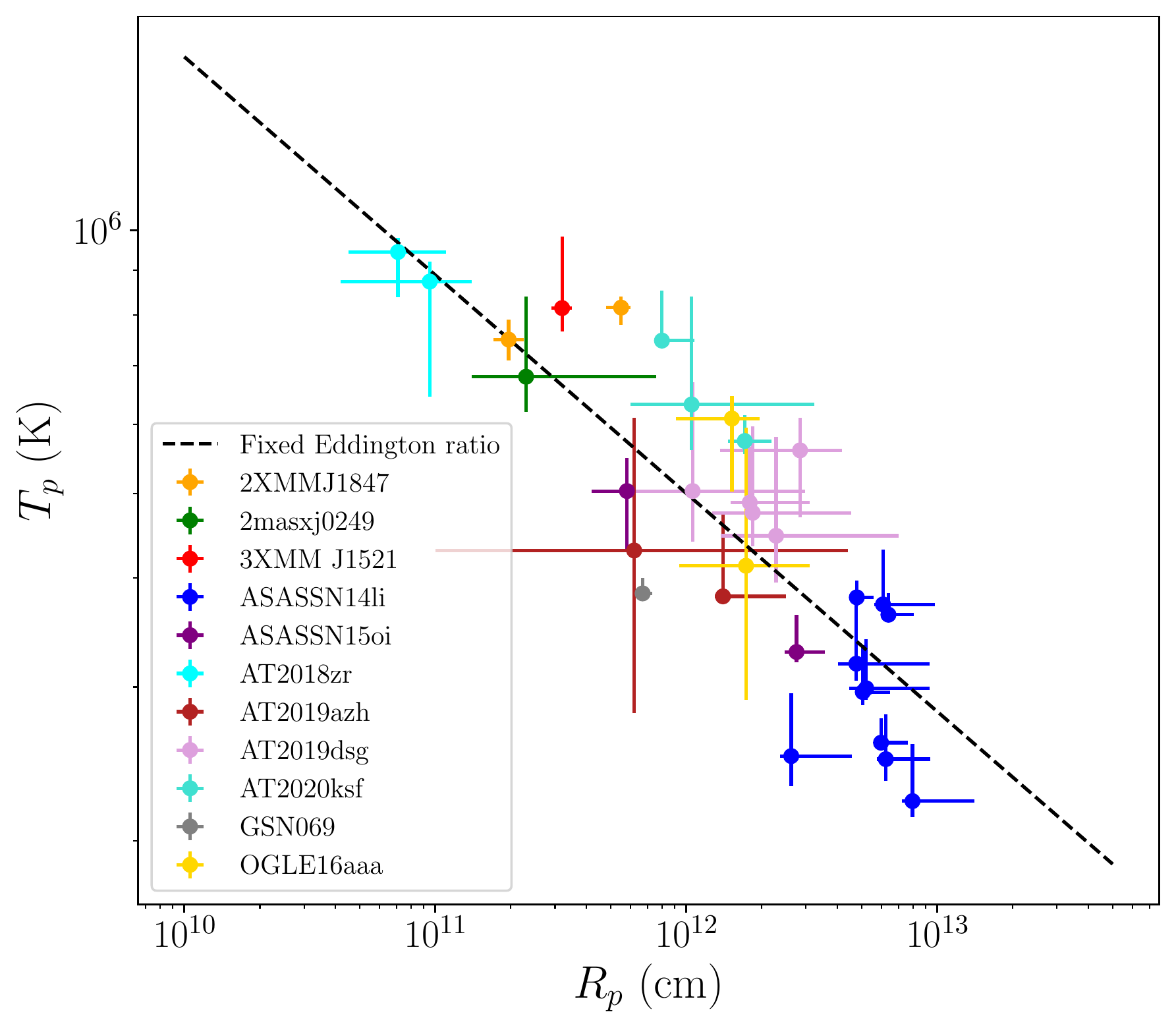}
    \caption{The inferred peak disc temperatures plotted against the radii at which this temperature occurred for all of the TDEs in our sample. We see a clear anti-correlation between temperature maximum and radius, which  follows a $T_p \propto R_p^{-1/4}$ relationship (black dashed curve). This is the exact relationship one would expect to find if the radius scaled linearly with the black hole mass, and the bolometric luminosity of these sources was a fixed fraction of the black hole Eddington luminosity. }
    \label{fig:fig3}
\end{figure}

\begin{figure}
    \centering
    \includegraphics[width=\linewidth]{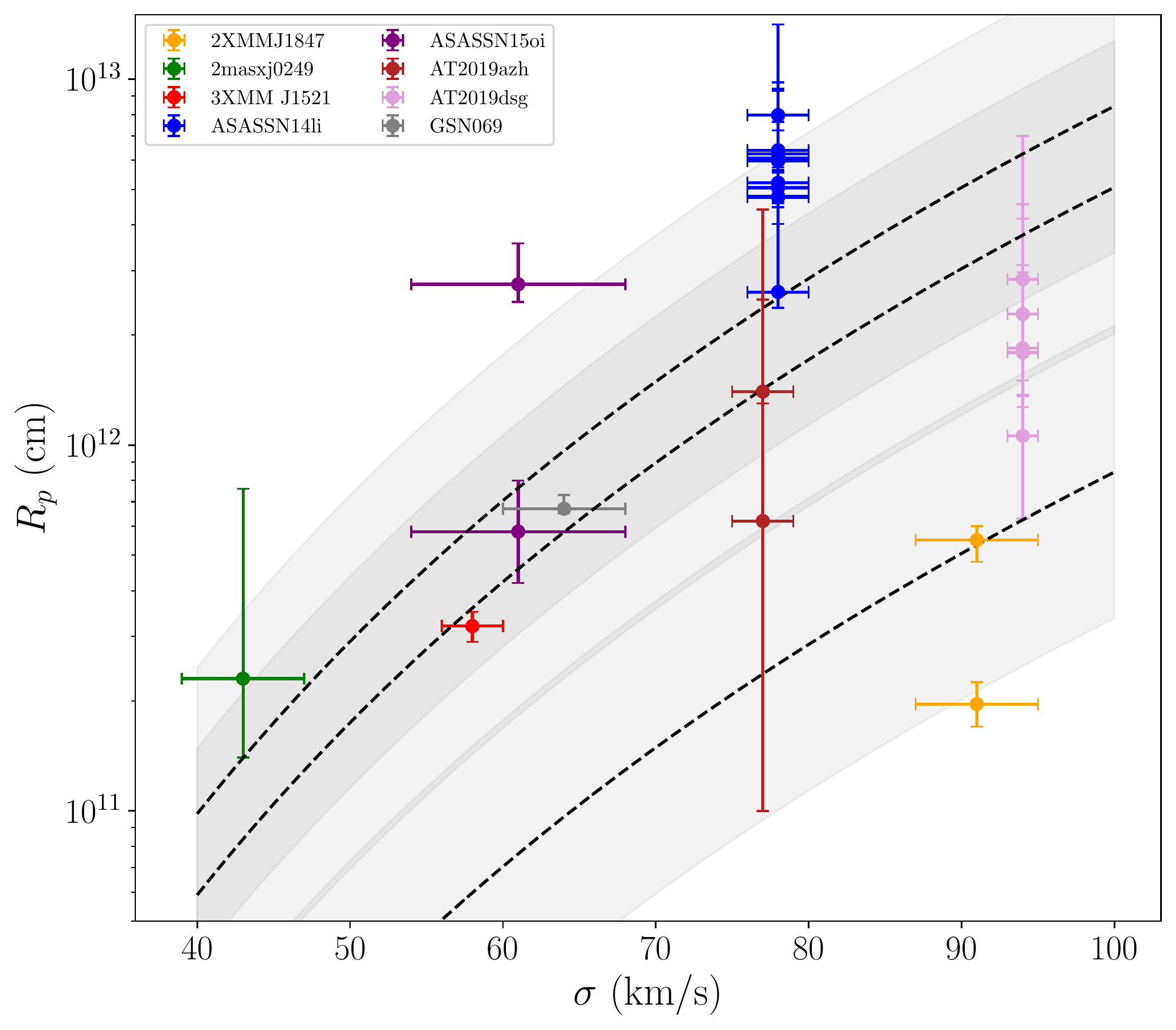}
    \caption{The radial location inferred from the X-ray spectra, plotted against the TDE host galaxies velocity dispersion $\sigma$. The dashed lines display the values of 1, 6 and 10 $GM_{\sigma}/c^2$, while the gray contours show the typical scatter in the $M-\sigma$ relationship (0.4 dex). The typical radii  inferred from TDE X-ray spectra lie in the parameter space which would be expected if these radii tracked the ISCO radius of the TDEs black hole.  }
    \label{fig:fig2}
\end{figure}

The parameter values inferred from each X-ray spectral fit are displayed in Table \ref{table:results}.  We find peak temperature values which range between $2 \times 10^5 < T_p ({\rm K}) < 1 \times 10^6$, and radii which span roughly two orders of magnitude $10^{11} < R_p ({\rm cm}) < 10^{13}$. This large radial range is not surprising given that our sample includes sources with velocity dispersion measurements ($\sigma$) which span a factor $\sim 2.5$. Super massive black hole masses are known to correlate strongly with velocity dispersion $M_{\rm BH} \sim \sigma^{\alpha}$, with $\alpha \sim 4.5-5.5$.

One interesting trend that is immediately obvious in our results is that the peak temperature of TDE discs are strongly anti-correlated with the inferred disc radii (Fig. \ref{fig:fig3}). Interestingly, the correlation approximately follows $T_p \propto R_p^{-1/4}$ (black dashed curve, Fig. \ref{fig:fig3}). This scaling is exactly what would be expected if the radius $R_p$ scaled linearly with black hole mass $R_p \propto M_{\rm BH}$, and the bolometric luminosity was a fixed fraction of the Eddington ratio $L_{\rm bol} \propto R_p^2 T_p^4 \propto L_{\rm edd} \propto M_{\rm BH}$. We reiterate that there is no intrinsic link between the temperature and radial parameters within the model itself, and that this correlation is a property of the sources themselves.  It appears therefore that there are two key physical correlations which deserve further attention: a trend between bolometric luminosity and central black hole mass, and between the X-ray radius and black hole mass. 

Those sources with both a high quality X-ray spectrum, and a measurement of the galaxy central velocity dispersion $\sigma$ are ideal targets for an analysis of any $R_p-M_{\rm BH}$ scaling relationship.  In Fig. \ref{fig:fig2} we plot the X-ray radial measurements against velocity dispersion for the 8 sources in our sample where both measurements are available. We display by dashed lines the values of 1, 6 and 10 $GM_{\sigma}/c^2$, with the mass values taken from the $M-\sigma$ relationship. The typical scatter ($\pm 0.4$ dex) in the $M-\sigma$ relationship is shown by the gray contours surrounding each curve. The typical radii  inferred from TDE X-ray spectra lie in the parameter space which would be expected if these radii tracked the ISCO radius of the black hole. This is in sharp contrast with the findings of previous works which relied on the use of a pure blackbody profile to describe the X-ray emission  (e.g. Brown et al. 2017, Holoien et al. 2018, van
Velzen et al. 2019, Wevers et al. 2019, Stein et al. 2020, Cannizzaro
et al. 2021, Hinkle et al. 2021), which typically find X-ray radii far smaller than the  gravitational radius of the host black hole. 

A simple polynomial fit of $R_p = A \sigma^B$, finds
\begin{equation}
    B = 2.1 \pm 1.4,
\end{equation}
which is somewhat shallower than the indices found for the $M_{\rm BH}-\sigma$ relationship (e.g., Ferrarese \& Ford 2005, McConnel \& Ma 2013). We do not suggest that this represents evidence for a break in the $M-\sigma$ relationship, due to our small sample size. However, future large samples of X-ray bright thermal state TDEs may represent the best probe of the $M-\sigma$ relationship at low $\sigma$, as TDEs preferentially occur around black holes with lower masses (e.g., Magorrian \& Tremaine 1999,  Stone \& Metzger 2016, Wevers et al. 2017).

In the following sub-sections we discuss a number of the implications of the results presented in Table \ref{table:results}. We begin with a discussion of the Eddington ratios of X-ray bright TDEs.

\subsection{The Eddington ratio of TDE discs}
It is well known that many accretion disc systems show dramatic changes in their behaviour as a function of the disc's Eddington ratio (Fender 2001,  Maccarone  2003, Maccarone et al. 2003, Fender \& Belloni 2004, Fender et al. 2004, Koerding, Jester \& Fender 2006, Ruan et al. 2019, Wevers 2020, Wevers et al. 2021). In particular, X-ray binary discs inhabit the so-called ``soft state'', spectrally analogous to the thermal TDE X-ray spectra studied in this paper, when their luminosity spans roughly $L_{\rm bol} \sim 0.01 - 1 L_{\rm edd}$.

With both the value and radial location of the peak TDE disc temperature determined from observations, the bolometric luminosity of the TDE disc can be self consistently estimated.   In a Newtonian theory of gravity, the bolometric disc luminosity is given by 
\begin{equation}
    L_{\rm bol} = \int^{R_{\rm out}}_{R_{\rm in}} 2\pi R\, 2\sigma_{SB} T^4(R) \,\,{\rm d}R ,
\end{equation}
where $R_{\rm in}$ and $R_{\rm out}$ are the inner and outer disc radii respectively, and $\sigma_{SB}$ is the Stefan-Boltzmann constant.  In a relativistic disc the only modification of this expression is the deviation from the local disc area element from $2\pi R {\rm d}R$. This minor simplification will not greatly affect our results.  

If we assume that there is minimal emission from interior to the radius at which the temperature peaks, and that the temperature falls exterior to this radius according to the classical $T(R) \propto R^{-3/4}$ accretion profile, we have 
\begin{equation}\label{LBOL}
    L_{\rm bol} = 4\pi\sigma_{SB} R_p^2 T_p^4 \left[ 1- \left(\frac{R_p}{R_{\rm out}}\right)\right] .
\end{equation}
It is important to note here that this luminosity  corresponds to the disc frame ``de-absorbed''  luminosity, as the effects of absorption of photons between the TDE and the observer have been modelled out with the inclusion of the {\tt TBabs} model (Table \ref{table:results}). We generally find low values for the absorbing hydrogen column density $n_H$ (Fig. \ref{fig:nsigma}), and so any errors introduced by incorrect modelling of the absorption are likely to be minimal. {We test the validity of the simplifications used in this modelling in Appendix C. We find that the errors associated with these simplifications are typically at the $\sim$ factor $2$ level, with equation \ref{LBOL} on average slightly overestimating the bolometric luminosity of the system. A factor $2$ uncertainty is less than the error range introduced by the uncertainty
in the fitted parameters for most observations in our sample (Table \ref{table:results}). }

\begin{figure}
    \centering
    \includegraphics[width=\linewidth]{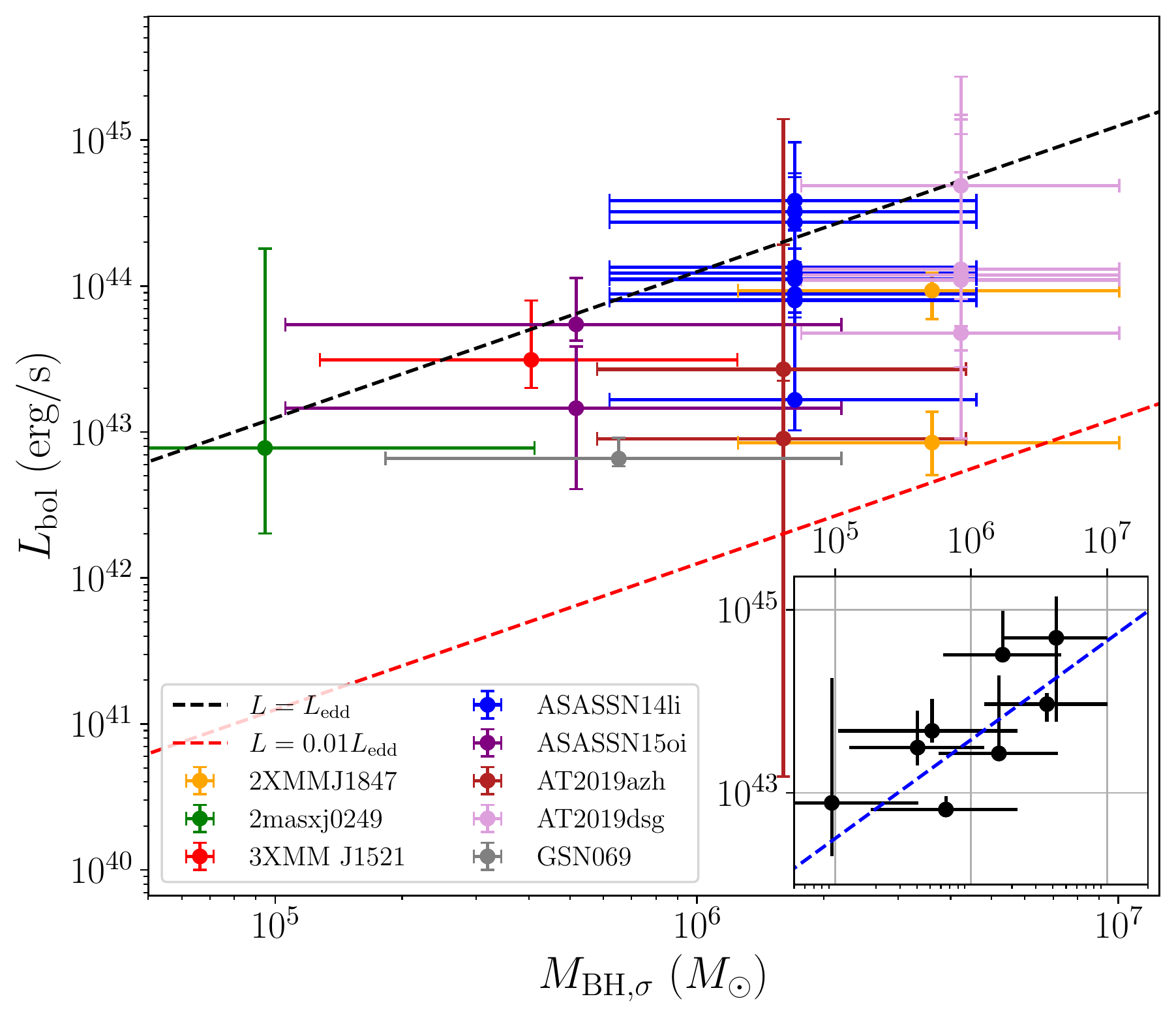}
    \caption{The inferred bolometric disc luminosity plotted against the black hole mass implied from the $M-\sigma$ relationship. Inset: the peak bolometric luminosity of each TDE plotted against $M-\sigma$ mass. We see a clear positive correlation between peak bolometric luminosity and black hole mass (best fit displayed by blue dashed curve in the inset). Each TDE observation is, within the error bars, consistent with being between $L = L_{\rm edd}$ (black dashed curve) and $L = 0.01 L_{\rm edd}$ (red dashed curve).  }
    \label{fig:bolmsig}
\end{figure}

We take $R_{\rm out}$ to be the circularisation radius of the TDE, whereby 
\begin{equation}
    R_{\rm out} = 94 \beta \, R_g \,  \left({M_{\rm BH} \over 10^6M_\odot}\right)^{-2/3}
\end{equation} 
for a solar type star. {In this expression $\beta$ is the ``penetration factor'' of the orbit (the ratio of the disrupted stars orbital pericentre  to the tidal radius $\beta = R_T/R_p \leq 1$). We will assume that $\beta = 1$ for the remainder of this paper. }This choice of outer radius has only a small effect on the inferred bolometric luminosity for our TDE sample\footnote{Choosing $R_{\rm out} \to \infty$ changes the individual luminosity values of our sample by no more than 15\% for any data point, and typically by $\sim 5$\%. On the other hand, if a TDE originated from a highly relativistic disruption with pericenter radius $R_{\rm peri} \sim R_I$, ($\beta \ll 1$) then the resulting bolometric luminosity may be smaller than the values given here. At late times the outer radius of all TDE discs will be large as a result of the conservation of angular momentum of the accreting disc material.  } which, as they generally have small inferred $M_{\rm BH, \sigma}$, have large outer radii (Figs. \ref{fig:bolmsig}, \ref{fig:fig1}). 

Using the values of $R_p$ and $T_p$ inferred from the TDEs X-ray spectra, and the black hole masses inferred from the $M_{\rm BH}-\sigma$ relationship, we plot the disc bolometric luminosity versus the $M_{\rm BH} - \sigma$ black hole mass in Figure \ref{fig:bolmsig}. The inset shows the peak bolometric luminosity observed from each source. The black dashed curve shows the Eddington luminosity
\begin{equation}
    L_{\rm edd} = 1\times 10^{45} \left({M_{\rm BH} \over 8 \times 10^6 M_\odot}\right) \,\, {\rm erg/s},
\end{equation}
while the red dashed curve is equal to $0.01 L_{\rm edd}$.

In Figure \ref{fig:bolmsig} we see a clear positive correlation between the peak bolometric luminosity and black hole mass of our TDE sample. A polynomial fit to the sample of $L_{\rm bol, peak} = A M_{\rm BH, \sigma}^B$ gives 
\begin{equation}
    B = 0.93 \pm 0.33,
\end{equation}
i.e., consistent with a linear relationship within the one standard deviation uncertainties, suggesting that TDE discs form at constant Eddington ratio. The best-fitting amplitude $A$ in this relationship corresponds to an Eddington ratio of $f_{\rm edd} = 0.37^{+0.46}_{-0.21}$. 

We further quantify this result by performing a power-law fit to these data, and taking into account the heteroscedastic measurement uncertainties. We use the Python implementation of the {\tt linmix} package (Kelly 2007) to perform linear regression in log-space:
\begin{equation}
    \log_{10}(L_{\rm bol, peak}) = \alpha + \beta  \log_{10}(M_{{\rm BH},\sigma}),
\end{equation}
{with $L_{\rm bol, peak}$ in units of erg s$^{-1}$, and $M_{{\rm BH},\sigma}$ in solar masses. }
To determine $\alpha$ and $\beta$ with their uncertainties, we bootstrap the data by uniformly sampling within the 1-$\sigma$ error bars and performing the correlation analysis on 1000 realizations. These results are consistent with a linear relationship, where the median and standard deviation are given by $\alpha = 37 \pm 2$ and $\beta = 1.1 \pm 0.3$ (see the inset of Figure \ref{fig:bolmsig}).

\begin{figure}
    \centering
    \includegraphics[width=\linewidth]{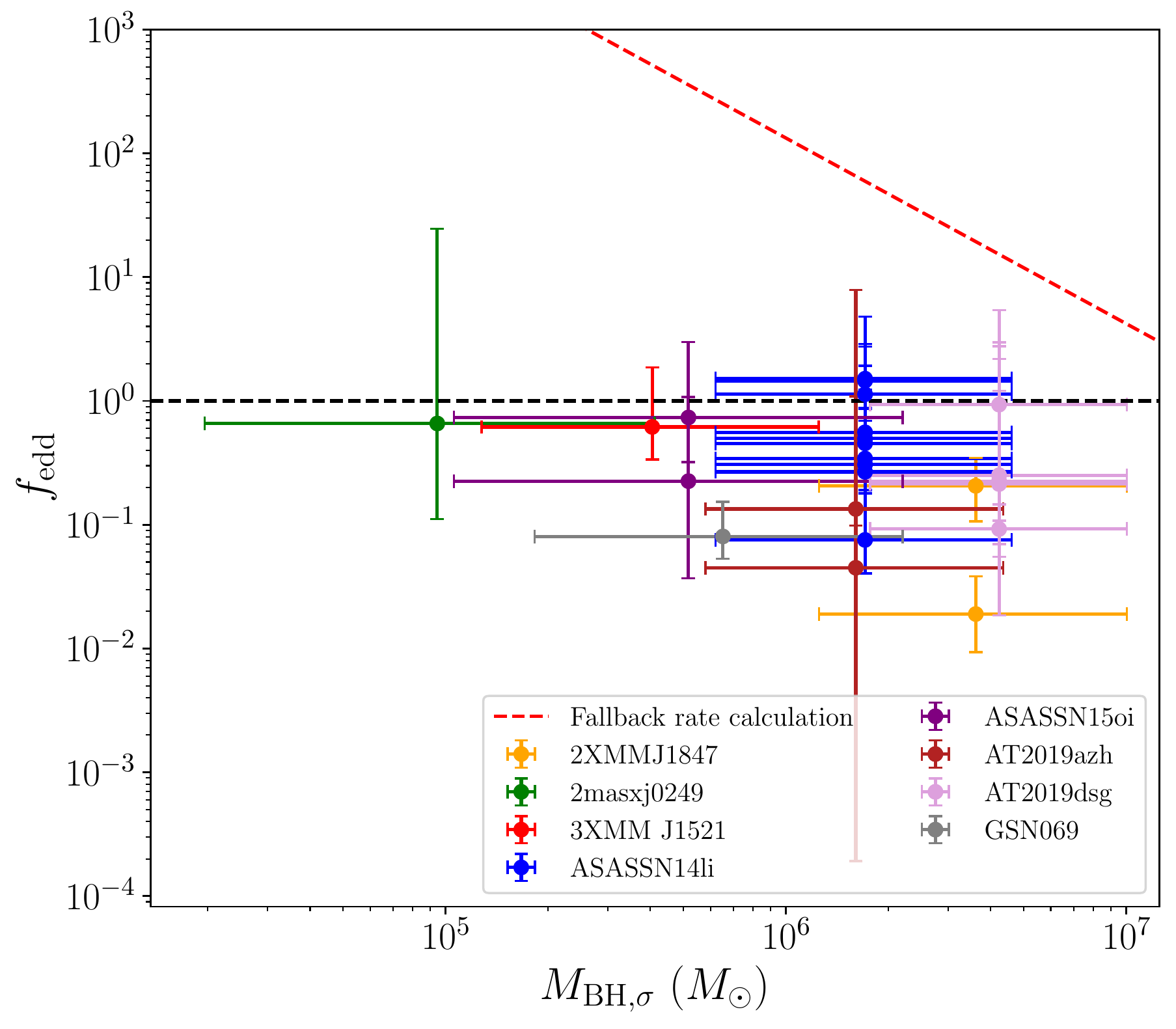}
    \caption{The Eddington {luminosity} ratio of the TDE sources with both a high quality X-ray spectrum and a galactic velocity dispersion measurement. The Eddington ratio is computed assuming that the TDE black hole mass is given by the $M_{\rm BH}-\sigma$ value, and that the bolometric luminosity is given by equation \ref{LBOL}. Every TDE is, within the uncertainties,  consistent with having a sub-Eddington luminosity $f_{\rm edd} \leq 1$. In addition, all TDE sources are, within the uncertainties,  consistent with having a luminosity higher than the hard-state transition luminosity seen in X-ray binaries $f_{\rm edd} \geq 0.01$.    }
    \label{fig:edrat}
\end{figure}

In Figure \ref{fig:bolmsig} we see that each TDE observation is, within the error bars, consistent with being between $0.01 L_{\rm edd} < L < L_{\rm edd}$. This can be seen more clearly in Figure \ref{fig:edrat}, where we plot the Eddington {luminosity} ratio of each TDE
\begin{equation}
    f_{\rm edd} \equiv L_{\rm bol}/L_{\rm edd} .
\end{equation}
It is interesting, but potentially not surprising, that the TDEs we have examined in this paper have Eddington ratios which are at the same level as those of  X-ray binaries in their ``soft'' state $0.01 \leq f_{\rm edd} \leq 1$. The TDEs that are modelled successfully by equation (\ref{MB}) are themselves in a thermal dominated  state. Our results suggests that the soft accretion state properties of black hole discs are universal, and in particular are black hole mass independent.

Finally, in Fig. \ref{fig:edrat}, we plot as a red dashed curve  the canonical `fallback rate' calculation of the Eddington {mass accretion rate} ratio $f_{\rm  fb} \equiv {\dot M_{\rm fb} / {\dot M}_{\rm edd}} \propto M_{\rm BH}^{-3/2}$ (Rees 1988), assuming a star of stellar mass and nominal accretion efficiency of $0.1$. We find no evidence for a fundamental link between the observed Eddington {luminosity} ratio of TDE discs and the fallback {accretion} rate {ratio}.  {For disc systems which are out  of inflow equilibrium, such as a TDE disc, the accretion rate does not represent a fundamental constant which encapsulates the physical properties of the disc. This is because the  accretion rate in these systems will vary at every disc radius, and also at each time.  It is likely that the bolometric luminosity of a TDE disc will better trace the mass accretion rate across its inner edge (at the ISCO), while the fallback rate represents the mass flux into the discs {\it outer} edge. The mass fluxes at the inner and outer edges of the disc are not equal and may in fact be very different (e.g., Mummery \& Balbus 2021). In addition, it is possible that winds launched in the early stages of TDE accretion remove material from the flow, restricting the luminosity to near Eddington values. 

Finally, steady-state discs at high mass accretion rates are known to deviate from a linear luminosity-accretion rate relationship (e.g., Abramowicz et al. 1998, Jiang et al. 2019), which may further explain why the fall-back rate fails to describe the observed luminosity evolution of our sources.   }

\subsection{(A lack of) Absorption in TDE X-ray spectra}
\begin{figure}
    \centering
    \includegraphics[width=\linewidth]{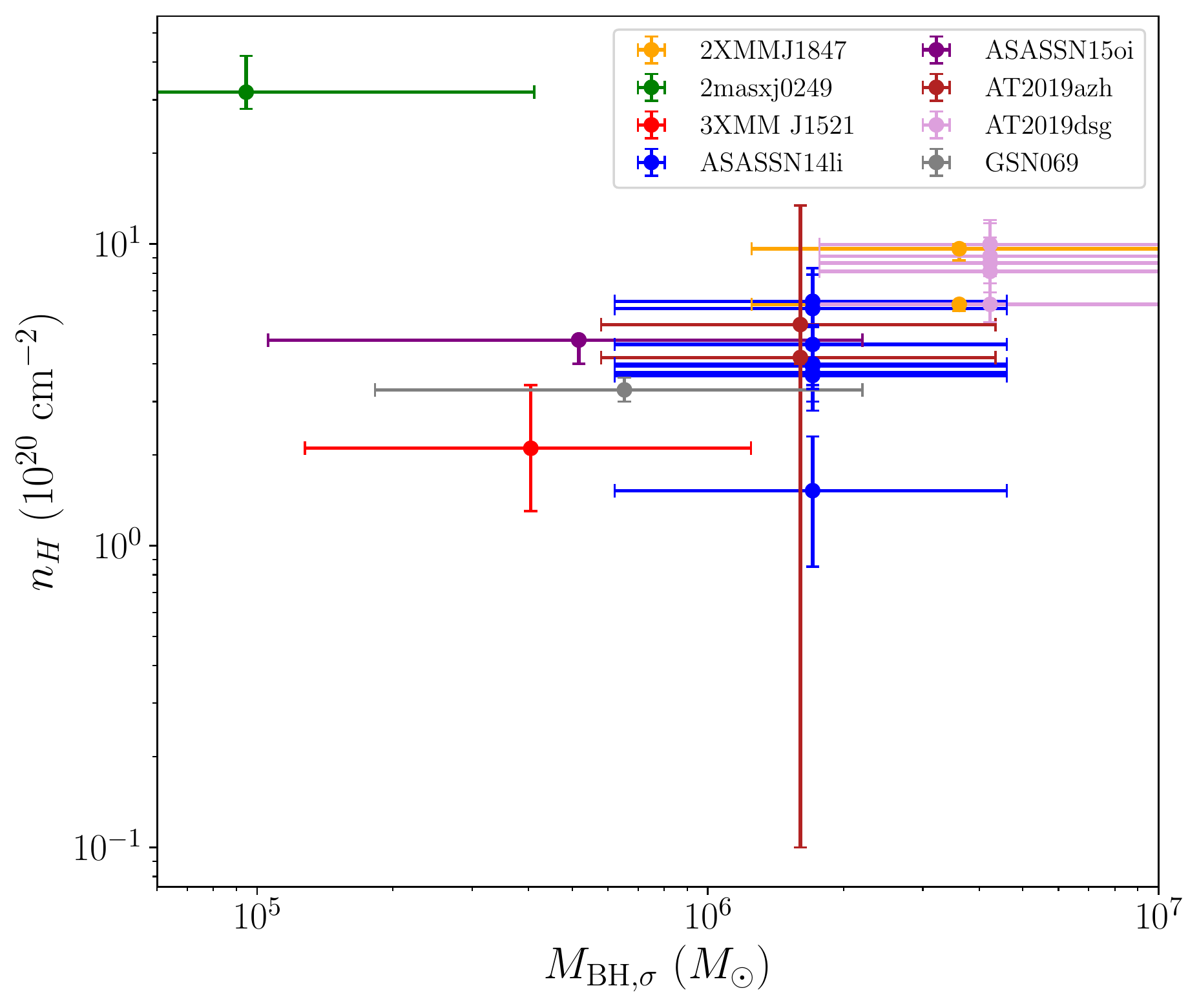}
    \caption{ The absorbing column depth of the {\tt TBabs} model, plotted against the $M_{\rm BH} - \sigma$ black hole mass, for the TDEs in our sample.  There is no correlation between $n_H$ and black hole mass, and all values of $n_H$ are relatively small $n_H \ll 10^{22}$ cm${}^{-2}$. Points displayed without vertical error bars are fixed at the galactic value for the column density.  }
    \label{fig:nsigma}
\end{figure}

\begin{figure}
    \centering
    \includegraphics[width=\linewidth]{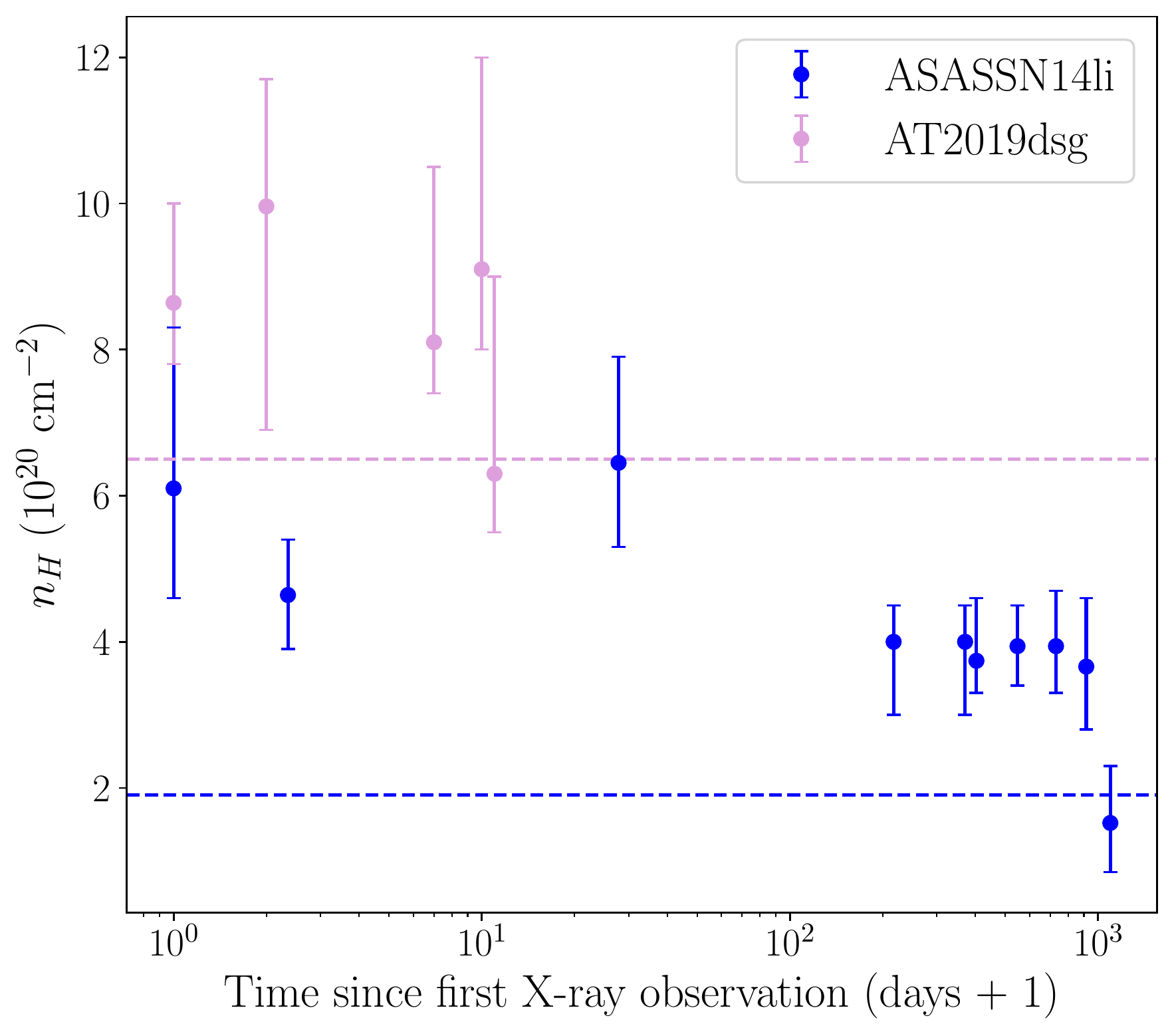}
    \caption{The evolving absorbing column depth of the {\tt TBabs} model, plotted against the time since the first X-ray observation of ASASSN-14li and AT2019dsg. The final observation of ASASSN-14li shows that the absorption has fallen to the level of the galactic column density in its direction (denoted by the dotted lines). }
    \label{fig:n_t}
\end{figure}

The results of the previous sub-section indicate that the bolometric luminosity of the TDEs in our sample are limited by the Eddington luminosity of their host black hole.  It is important to verify that this result is not produced by an increasing absorbing column density at low black hole masses, which could potentially act as the dominant  systematic effect in our sample.  

In Figure \ref{fig:nsigma} we plot the absorbing column density $n_H$ from the {\tt TBabs} model against the $M_{\rm BH}-\sigma$ black hole mass of the TDEs in our sample.  There is no correlation between $n_H$ and black hole mass, and all values of $n_H$ are relatively small\footnote{The one potential exception to this finding is 2MASXJ0249, which has a moderate absorbing column density $n_H \simeq 3 \times 10^{21}$ cm${}^{-2}$. We note that the detailed spectral modelling of Strotjohann {\it et al}. (2016) found that 2MASXJ0249 was best described by a complex absorber with both cold and ionised gas. } $n_H < 10^{21}$ cm${}^{-2}$. A systematic effect resulting from an increase in $n_H$ around low black hole mass TDEs can be ruled out as driving the $L_{\rm bol}-M_{\rm BH}$ correlation of the previous sub-section. This result is important but potentially unsurprising,  as we have selected a sample of bright thermal X-ray sources which must will by construction have modest absorbing column depths.

It is interesting to examine the evolution of the absorbing column density for the two TDE's where we have the best temporal coverage, ASASSN-14li and AT2019dsg (Figure \ref{fig:n_t}).  Both sources are observed to have absorbing column densities above the galactic level (denoted by dotted lines) at early times. At the latest times it appears that, particularly for ASASSN-14li, the absorbing column density of the sources evolves to the galactic level (as also found by Kara et al. 2018,  Wen et al. 2020).  One interpretation of this behaviour would be a wind of material which is launched in the early stages of the TDE, which at early times acts as an absorbing column, but which is not present, or is significantly weaker, at late times. 

\subsection{Measuring black hole masses from TDE X-ray spectra}

The correlation between $R_p$ and $\sigma$ (Figure \ref{fig:fig2}) suggests that the X-ray radius $R_p$ can be used as a measurement of the black hole mass. 
This is unsurprising, as the parameter $R_p$ corresponds to the observed location at which the TDE disc temperature is a maximum.  As this radial co-ordinate will likely correspond to the region around the ISCO (Mummery 2021), it will scale linearly with the physical radius of the TDE's accretion disc.   As highlighted in Mummery (2021) however, there are two compounding parameter degeneracies which must be taken into account before this inversion can be made.  The first relates to the inclination angle between the TDE disc and the observer, which modifies the projected image plane radius of the temperature region. The second intrinsic degeneracy between radius and mass relates to the black hole's spin, which changes the size of the ISCO in units of the black hole's mass.  

To examine the effects of varying black hole spin and disc-observer inclination angle on the inferred radial size parameter $R_p$, we numerically simulate and fit a number of 0.3--2 keV X-ray spectra with known black hole masses, spins and inclinations with the model of equation \ref{MB}. The mock X-ray spectra are generated by solving the relativistic thin disc equations, and ray-tracing the resulting temperature profiles (see e.g., Mummery \& Balbus 2020 for a description of the algorithms used to solve both the disc evolution and photon orbit equations). The mock spectra were then fit with equation \ref{MB}, and the parameter $R_p$ was compared to the (known) value of both the ISCO radius $R_I$ and the (known) gravitational radius $R_g \equiv GM_{\rm BH}/c^2$. 

\begin{figure}
    \centering
    \includegraphics[width=\linewidth]{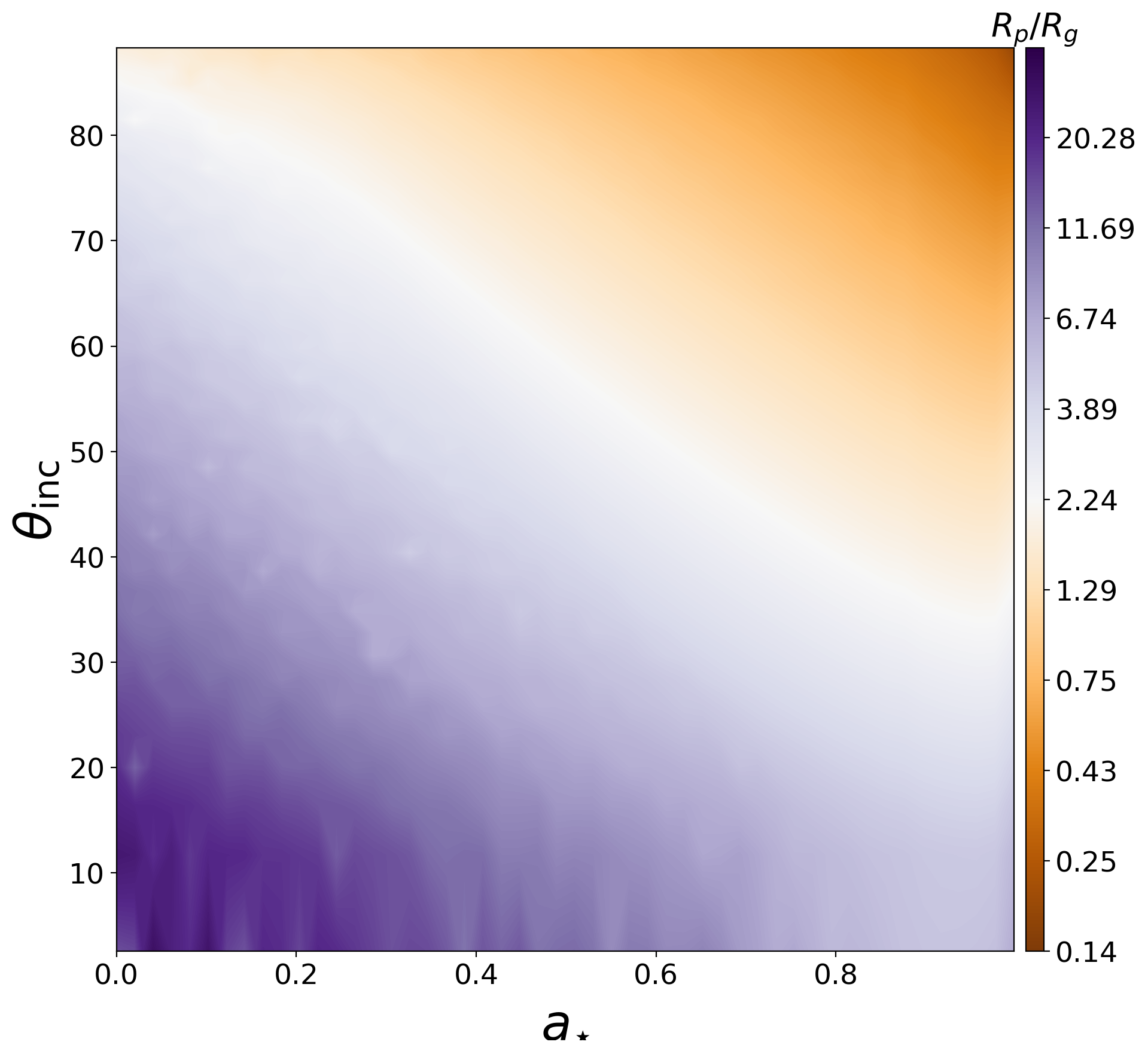}
    \caption{The distribution of $Y = c^2 R_p/GM_{\rm BH}$ as a function of black hole spin and inclination angle. Increasing either the spin or inclination angle causes the inferred $R_p$ parameter to decrease.    }
    \label{fig:Rc}
\end{figure}

\begin{figure}
    \centering
    \includegraphics[width=\linewidth]{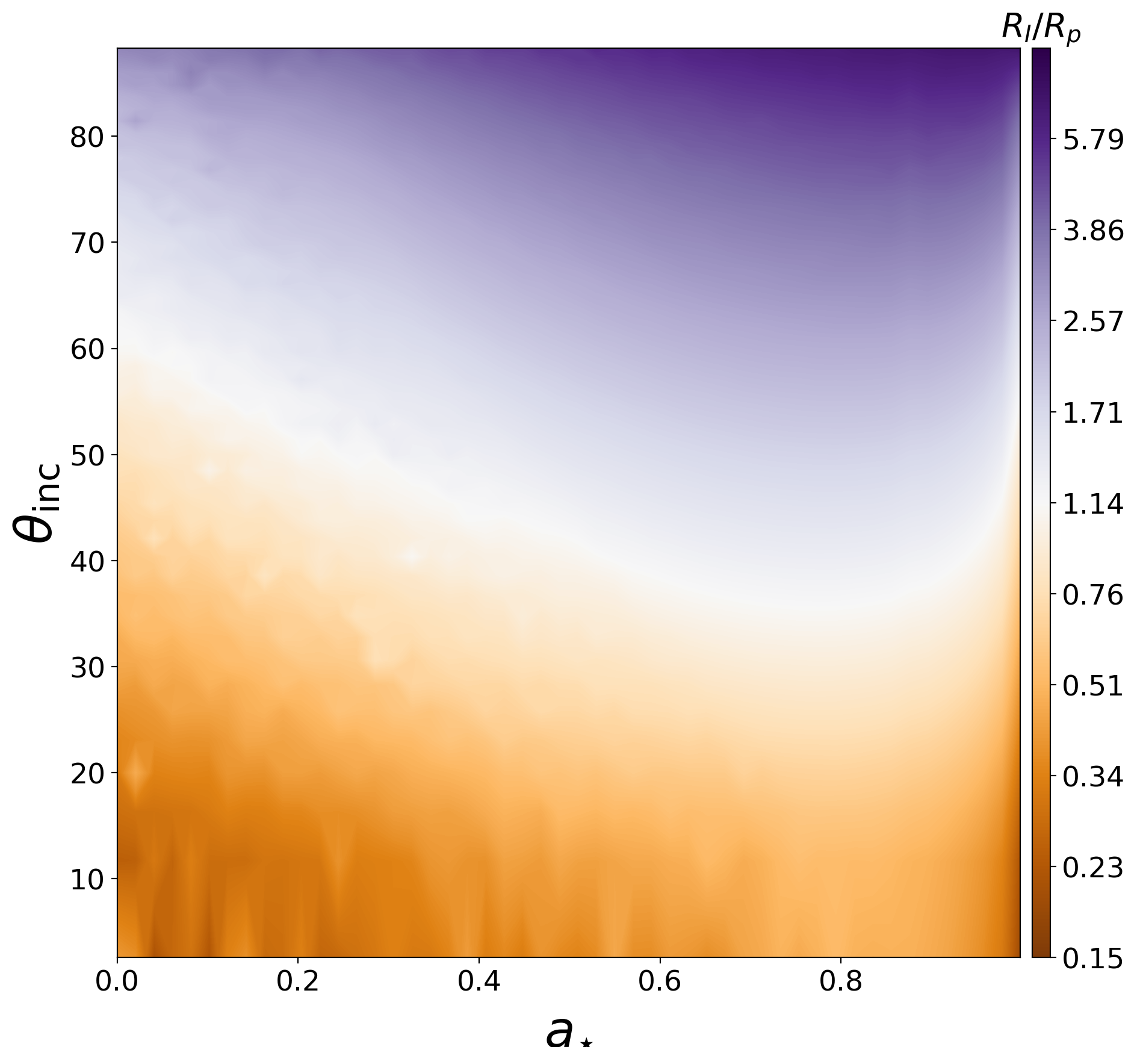}
    \caption{The distribution of the ratio $R_I/R_p$ as a function of black hole spin and inclination angle. While there remains some trend with spin, this ratio is principally dependent on $\theta$. The white line at approximately $\theta \simeq 50^\circ$ shows that the model performs best at this inclination.  }
    \label{fig:DR}
\end{figure}

In Figure \ref{fig:Rc} we display the inferred $R_p$ parameter, in units of the gravitation radius $R_g$ (which we denote by $Y \equiv R_p/R_g$) as a function of black hole spin and inclination angle. Typically the radius inferred from the spectral observations lies in the range of 1 to 10 gravitational radii, with extremes a factor two higher (lower) for low (high) spins and low (high) inclinations. The ratio of $R_p$ to the location of the ISCO radius is shown in Figure \ref{fig:DR}, indicating that a measurement of $R_p$ should provide a good estimate of the size of the TDE disc's ISCO for most inclination angles, with large errors only at high $\theta_{\rm inc} > 80^\circ$ and low $\theta_{\rm inc} < 10^\circ$ inclinations. The ratio $R_I/R_p$ is principally only dependent on the inclination of the disc, with higher inclinations leading to $R_p < R_I$, and smaller inclinations $R_p > R_I$. While the unknown value of $\theta_{\rm inc}$ introduces scatter into the $R_p-M_{\rm BH}$ relationship, if the TDE black hole spins are assumed to be uniformly distributed, and the observed inclination angles are assumed to be distributed uniformly on a sphere, then the mean value of the ratio $R_I/R_p$ across the whole population is found to be $0.97$.

\begin{figure}
    \centering
    \includegraphics[width=\linewidth]{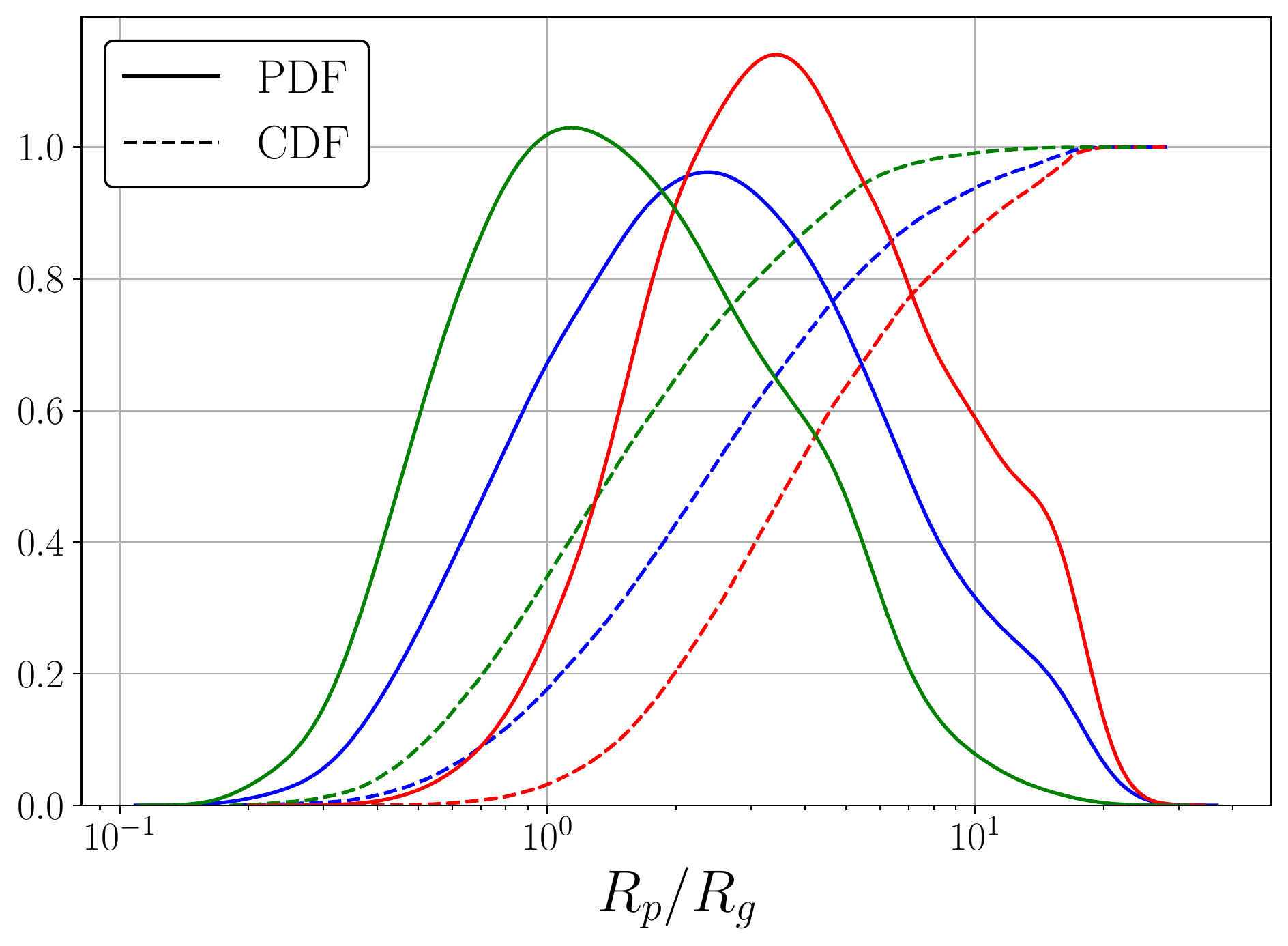}
    \caption{Probability density function (PDF, solid curves) and cumulative distribution function (CDF, dashed curves) of $Y = c^2 R_p/GM_{\rm BH}$, for three different assumptions of the underlying SMBH black hole spin distribution. The blue curves are for a uniform spin distribution ($p(a) = 1$), the green curves are for a spin distribution $p(a) \propto a^2$ (i.e., higher spins are favoured), while the red curves are for a spin distribution $p(a) \propto (a-1)^2$ (i.e., lower spins are favoured).   }
    \label{fig:Ydf}
\end{figure}

The distribution of the black hole masses one will infer from the radial parameter $R_p$ for a population of TDEs will depend on the intrinsic distributions of both the inclination angle $\theta$ and the black hole spins $a$. It is likely that TDE sources will be distributed randomly on a sphere, meaning that $\cos\theta$ will be uniformly distributed.  The distribution of TDE black hole spin's however is much less certain.  In Fig. \ref{fig:Ydf} we plot the probability density functions (PDFs, solid curves) and cumulative distribution functions (CDFs, dashed curves) of the variable $Y = R_p/R_g$ (i.e., the radius inferred from the X-ray spectra in units of the black hole mass) for three different intrinsic TDE spin  distributions. In blue, the spin distribution is assumed to be uniform, while the green curves are for a spin distribution\footnote{The notation $p(a)$ here denotes the probability density function of the black hole spin. } $p(a) \propto a^2$ (i.e., higher spins are favoured), while the red curves are for a spin distribution $p(a) \propto (a-1)^2$ (i.e., lower spins are favoured). While the choice of the black hole spin distributions does affect the black hole mass estimate, within the uncertainties inherent to this sort of modelling (the systematic offsets between the different distributions are smaller than their widths), it is unlikely to affect the interpretation of our results.  

\begin{figure}
    \centering
    \includegraphics[width=\linewidth]{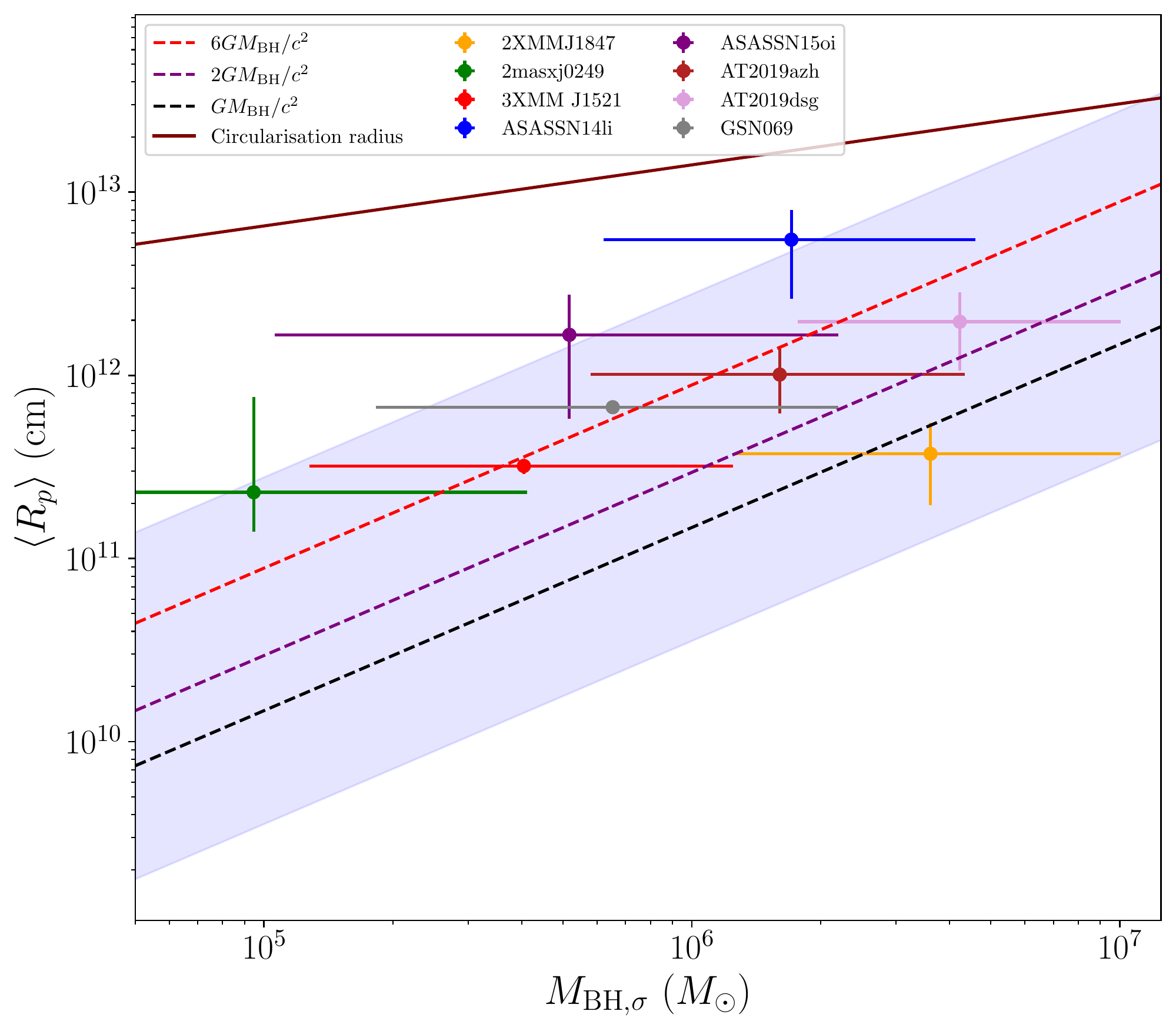}
    \caption{The average of the peak-temperature radial location inferred from the X-ray spectra, plotted against the black hole mass implied by the $M-\sigma$ relationship. Plotted as a brown solid curve is the circularisation radius of a TDE, expected to be the disc {\it outer} edge. Shown as dashed curves are the values 1, 2 and 6 $GM/c^2$. The blue shaded region shows 99\% confidence region of the uniform black hole spin distribution (Fig. \ref{fig:Ydf}).  The TDE sources studied in this paper occupy the expected region of $R_p-M_{\rm BH, \sigma}$ parameter space.    }
    \label{fig:fig1}
\end{figure}

In Fig. \ref{fig:fig1} we plot the mean radius inferred from the X-ray spectral measurements of each TDE against the black hole masses inferred from the $M-\sigma$ relationship, using our measurements of the velocity dispersion (Table \ref{table:sample}). Shown as dashed curves are the values 1, 2 and 6 $GM/c^2$. The blue shaded region shows 99\% confidence region of the uniform black hole spin distribution (Fig. \ref{fig:Ydf}). 

As is clear in Fig. \ref{fig:fig1}, despite the large scatter inherent to the $M-\sigma$ relationship, and the uncertainty in our measurements in $R_p$, the TDE sources studied in this paper occupy the expected region of $R_p-M_{\rm BH, \sigma}$ parameter space.

Assuming a uniform spin distribution, we can calibrate a radius-to-mass conversion factor $X$, defined as 
\begin{equation}
    \left( { M_{{\rm BH}, R_p} \over 10^6 M_\odot}\right) = X  \left({R_p \over 10^{12} \, {\rm cm}}\right)   .
\end{equation}
We find a mean value of $X$ of 
\begin{equation}\label{conversion_factor}
    \overline{X} \simeq 4.9^{+7.1}_{-3.0},
\end{equation}
{Where the error range denotes the 1$\sigma$ confidence interval (note that this interval corresponds to roughly $\pm0.4$ dex).} From this conversion  we can compute an X-ray spectral fit black hole mass for each TDE. Note that this particular of value $\overline{X}$ implies that $R_p \simeq 1.4 R_g$, and that a radius $R_p = 2 \times 10^{11}$ cm corresponds to a black hole mass of $\sim 10^6 M_\odot$ (i.e., all sources bar AT2018zr studied in this work are consistent with having a black hole mass $M>10^6 M_\odot$).

In Table \ref{tab:mass} we collate the black hole masses inferred from the mean X-ray radius of the TDEs examined in this paper. As a measure of the uncertainty in this conversion we include error ranges which correspond to the 5\%-95\% confidence region of the radius-to-mass conversion distribution, assuming a uniform distribution of black hole spins (blue curve, Figure. \ref{fig:Ydf}). We note that two of these sources, ASASSN-14li and ASASSN-15oi, have had black hole masses inferred from their X-ray spectra using relativistic slim disc modelling (Wen et al. 2020). It is encouraging that our inferred best-fitting masses ($M_{14{\rm li}} = 10 \times 10^6 M_\odot$, $M_{15{\rm oi}} = 5 \times 10^6 M_\odot$) are both consistent with the values found from the more complex models of Wen et al. (2020): $M_{14{\rm li}} = 10 \times 10^6 M_\odot$, $M_{15{\rm oi}} = 4 \times 10^6 M_\odot$, albeit with wider uncertainties.

\begin{table}
    \renewcommand{\arraystretch}{1.5}
    \centering
    \begin{tabular}{c|c}
       Source & $\log_{10} M_{\rm BH}/M_\odot$ \\
        \hline
        ASASSN-14li & $7.19^{+0.62}_{-0.66}$ \\ \hline 
        ASASSN-15oi & $6.67^{+0.63}_{-0.66}$ \\ \hline
        OGLE16aaa & $6.66^{+0.62}_{-0.66}$ \\ \hline
        AT2018zr & $5.37^{+0.62}_{-0.66}$ \\ \hline 
        AT2019azh & $6.46^{+0.62}_{-0.66}$ \\ \hline
        AT2019dsg & $6.74^{+0.63}_{-0.66}$ \\ \hline 
        AT2020ksf & $6.53^{+0.62}_{-0.67}$\\ \hline
        GSN069 & $6.28^{+0.62}_{-0.66}$ \\ \hline
        2XMM J1847 & $6.02^{+0.62}_{-0.66}$ \\ \hline
        2MASX J0249 & $5.81^{+0.62}_{-0.66}$\\ \hline
        3XMM J1521 & $5.96^{+0.62}_{-0.66}$\\ \hline 
         
    \end{tabular}
    \caption{The black hole masses inferred from the mean X-ray radius of the TDEs examined in this paper. The error ranges correspond to the 5\%-95\% confidence region of the radius-to-mass conversion distribution, assuming a uniform distribution of black hole spins (blue curve, Figure. \ref{fig:Ydf}). The values themselves correspond to the 50th percentile of the distribution.   }
    \label{tab:mass}
\end{table}

{The inversion procedure developed here has relatively large intrinsic scatter, of order $\sim 0.6$ dex. It may be possible to shrink this uncertainty further by incorporating additional spectral information (the measured disc temperature) into the parameter inversion procedure. This has promise as the disc temperature is also found to scale with the disc radius (Fig. \ref{fig:fig3}), a result of the near-universal Eddington luminosity ratio found in this work, and this may be an interesting future extension of this analysis.  

However, the fact that a given TDE source will be accreting with a near-Eddington luminosity is not guaranteed to be true apriori, and it seems reasonable to assume that sources with $f_{\rm edd} \neq 1$ will exist at some non-zero rate in the total TDE population. The time-dependent cooling of the disc temperature may also complicate the inversion procedure, and we therefore do not incorporate the disc temperature into the analysis performed here.  
}

\begin{figure}
    \centering
    \includegraphics[width=\linewidth]{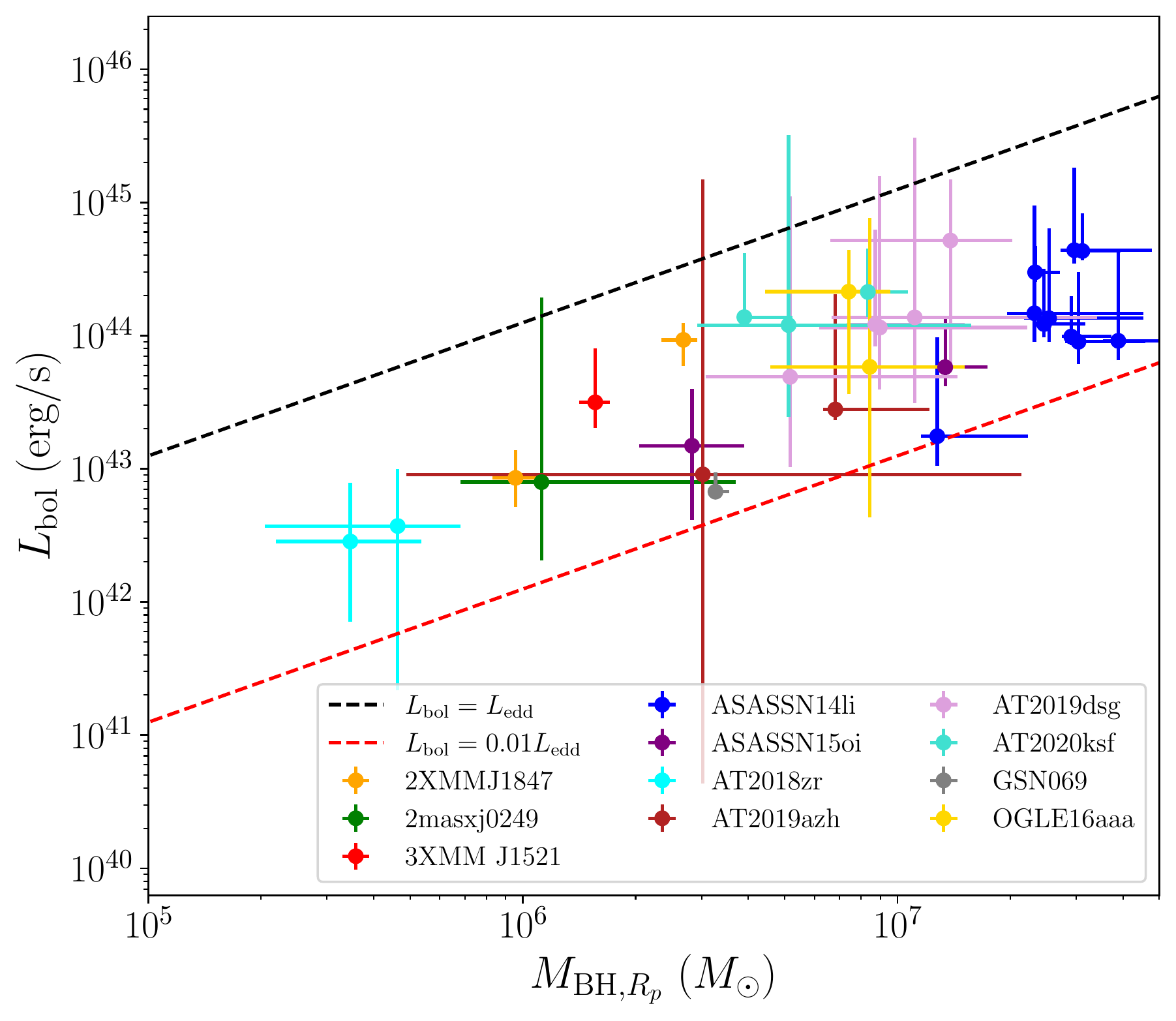}
    \caption{The inferred bolometric disc luminosity plotted against the $M_{\rm BH}-R_p$ mass for all of the TDEs in our sample. We see a clear positive correlation between bolometric luminosity and black hole mass. This is the exact relationship one would expect to find if the bolometric luminosity of these sources was a fixed fraction of the black hole's Eddington luminosity. }
    \label{fig:bolrad}
\end{figure}

\begin{figure}
    \centering
    \includegraphics[width=\linewidth]{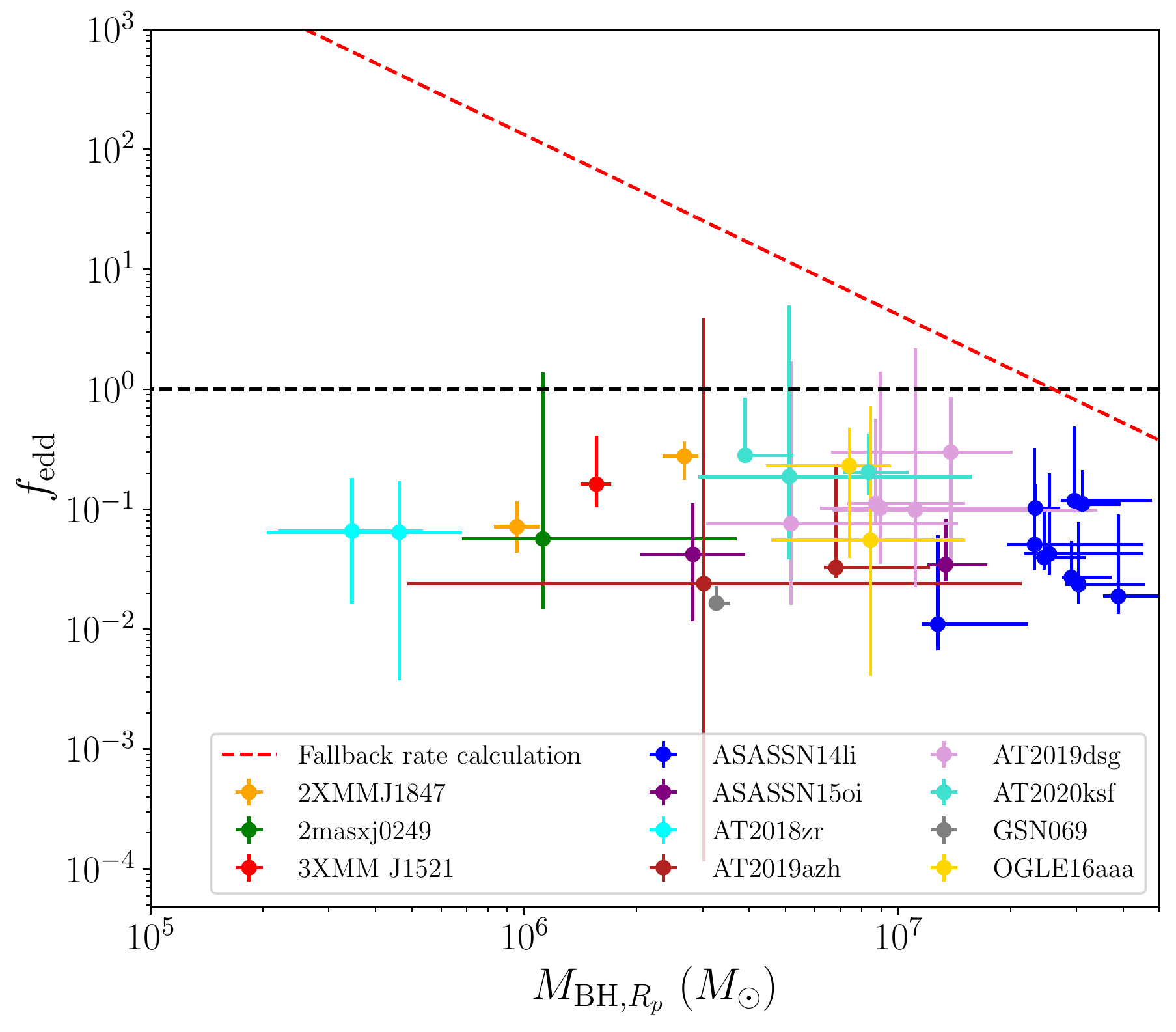}
    \caption{The Eddington {luminosity} ratio of the TDE sources in our sample. The Eddington ratio is computed assuming that the TDE black hole mass is given by the $M-R_p$ conversion of equation \ref{conversion_factor}, and that the bolometric luminosity is given by equation \ref{LBOL}. Every TDE is, within the uncertainties,  consistent with having a sub-Eddington luminosity $f_{\rm edd} \leq 1$. In addition, all TDE sources are, within the uncertainties,  consistent with having a luminosity higher than the hard-state transition scale seen in X-ray binaries $f_{\rm edd} \geq 0.01$.    }
    \label{fig:edratRP}
\end{figure}

\begin{figure}
    \centering
    \includegraphics[width=\linewidth]{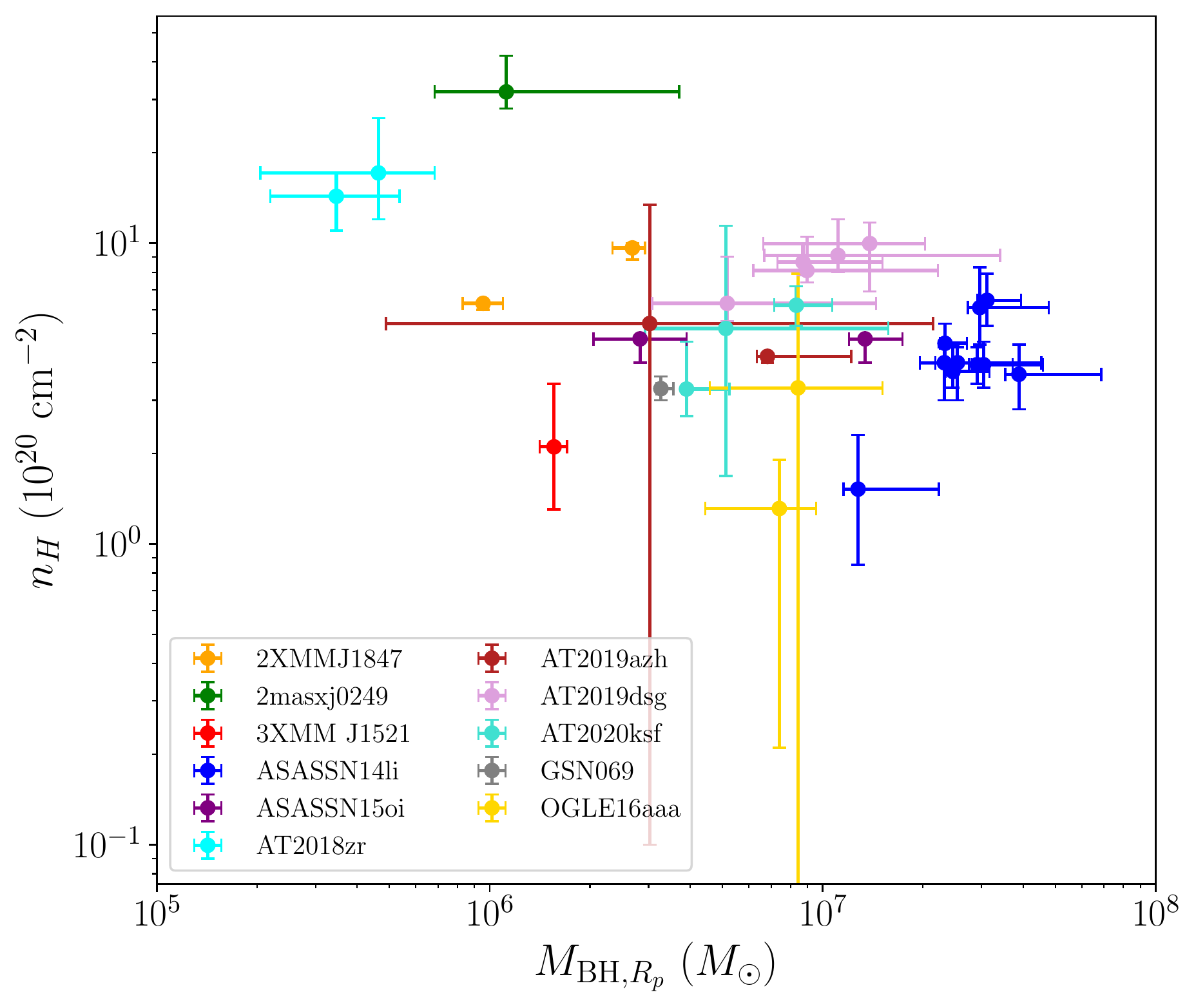}
    \caption{  The neutral absorbing column depth of the {\tt TBabs} model, plotted against the $M_{\rm BH} - R_p$ black hole mass, for the TDEs in our sample.  There is no correlation between $n_H$ and black hole mass, and all values of $n_H$ are relatively small $n_H \ll 10^{22}$ cm${}^{-2}$. Points displayed without vertical error bars are fixed at the galactic value for the column density.   }
    \label{fig:nr}
\end{figure}

With this radius-to-mass conversion factor determined, we plot in Figures \ref{fig:bolrad}, \ref{fig:edratRP} and \ref{fig:nr} the bolometric luminosity, Eddington ratio and absorbing column depth against $M_{{\rm BH}, R_p}$, the black hole mass determined from each X-ray radius. We again find that the bolometric luminosity scales approximately linearly with black hole mass, and that the Eddington ratios of these discs is limited to the range $0.01 \leq f_{\rm edd} \leq 1$. Note that the values of the luminosities and Eddington ratios of the sources with $\sigma$ measurements in Figs. \ref{fig:bolrad} and \ref{fig:edratRP} differ slightly from those of Figs. \ref{fig:bolmsig} and \ref{fig:edrat}. This is due to the use of $M_{\rm BH} - R_p$ black hole mass values in calculating $L_{\rm bol}$ and $f_{\rm edd}$, not the $M_{\rm BH} - \sigma$ values.   In addition, once again, no relationship between absorbing column density and black hole mass is found.

This result  indicates that the finding of Eddington limited TDE accretion is robust, and does not depend on use of the $M_{\rm BH}-\sigma$ relationship.

\subsection{The missing energy problem} 
The integrated observed energy in TDE X-ray light curves typically reaches values of $E_X \sim 10^{50}$ erg (e.g., Holoien et al. 2014b), far below the energies expected from the accretion of a significant fraction of the incoming stellar mass  $M_{\rm acc} \sim 0.1 M_\odot$ ($E_{\rm rad} \sim 10^{52}$ erg). By modelling the X-ray and UV light curves of a number of TDEs, Mummery (2021b) argued that the discrepancy between the observed and expected radiated energies could be explained by the large ($\eta_X \sim 10-100$) bolometric corrections inherent to TDE discs, a line of reasoning also discussed by Saxton {\it et al}. (2021). This argument can be further tested with the results of our X-ray spectral fitting. {We will now demonstrate analytically that the observed X-ray luminosity of a typical TDE disc will be an exponentially small fraction of the total disc luminosity, and that this fraction further decreases exponentially as a function of time (i.e., as the disc cools one observes a smaller and smaller fraction of the total energy emitted in the X-rays).  }

The {observed} X-ray luminosity of the accretion disc is given by the integral of the disc spectrum (eq. \ref{MB}) over the X-ray bandpass of the telescope\footnote{This calculation explicitly assumes that the disc remains in a thermal state independent of the disc temperature. This is unlikely to be universally valid: at the very lowest temperatures TDE disc's will likely transition into a harder state, where accretion energy is diverted into creating hot coronal electrons. This will increase $L_X$ and $\eta_X$ will subsequently reduce.  }:
\begin{equation}
    L_X = 4\pi D^2\int_{\nu_l}^{\nu_u} F_\nu(\nu, R_p, T_p, \gamma, D) \,\,{\rm d}\nu, 
\end{equation}
where $E_l = 0.3$ keV, and $E_u = 10$ keV. Once again, this value corresponds to the disc-frame ``de-absorbed'' luminosity. Combined with the bolometric luminosities of the previous sub-section, we can calculate another key parameter of the TDE system: the bolometric correction $\eta_X$, defined as
\begin{equation}
    \eta_X \equiv L_{\rm bol}/L_X .
\end{equation}
This bolometric correction quantifies the fraction of the total energy liberated from the accretion disc which is observed at X-ray frequencies.

As both the X-ray and bolometric disc luminosities have leading dependencies which scale as $R_p^2$, the bolometric correction only depends on the peak disc temperature $T_p$ and $\gamma$ parameter\footnote{Formally the bolometric correction also depends on the disc's {outer} radius, but this effect is minimal (of order a few percent for typical TDE black hole masses). }. An analytical estimate of the bolometric correction can be obtained by approximating  
\begin{equation}
    L_{\rm bol} \simeq 4\pi \sigma_{SB} R_p^2 T_p^4,
\end{equation}
and, using equation \ref{MB} (extending the upper integration limit to $+\infty$ introduces exponentially small corrections): 
\begin{multline}
        L_X \simeq {16 \pi^2 R_p^2 \xi_1 h \over c^2 f_{\rm col}^4}  \int_{\nu_l}^\infty \nu^3 \\ \left[ \left(\frac{k \T_p}{h \nu} \right)^\gamma + \xi_2\left(\frac{k \T_p}{h \nu} \right)^{1+\gamma} + \xi_3\left(\frac{k \T_p}{h \nu} \right)^{2 + \gamma}   \right] \exp\left(- \frac{h\nu}{k \T_p} \right)  {\rm d}\nu . 
\end{multline}
The solution to this integral can be written in terms of incomplete $\Gamma$ functions, defined as:
\begin{equation}
    \Gamma(s, z) \equiv \int_z^\infty t^{s-1} \exp(-t) \, {\rm d}t .
\end{equation}
Explicitly, after defining $t \equiv h\nu/k\T_p$, we have 
\begin{multline}
    L_X \simeq {16 \pi^2 R_p^2 \xi_1 h \over c^2 f_{\rm col}^4}  \left({k\T_p \over h}\right)^4 \\ 
    \int_{h\nu_l/k\T_p}^\infty  \left[ t^{3 - \gamma} + \xi_2 t^{2-\gamma} + \xi_3 t^{1 - \gamma}   \right] \exp\left(- t \right)  {\rm d}t ,
\end{multline}
with solution
\begin{multline}
        L_X = {16 \pi^2 R_p^2 \xi_1 h \over c^2 f_{\rm col}^4}  \left({k\T_p \over h}\right)^4 \\ \left[\Gamma\left(4 - \gamma, {h\nu_l \over k\T_p}\right) + \xi_2 \Gamma\left(3 - \gamma, {h\nu_l \over k\T_p}\right) + \xi_3 \Gamma\left(2 - \gamma, {h\nu_l \over k\T_p} \right) \right]. 
\end{multline}
Therefore 
\begin{multline}\label{etax}
    \eta_X \simeq {2\pi^4 \over 15 \xi_1} \Bigg[\Gamma\left(4 - \gamma, {h\nu_l \over k\T_p}\right) + \xi_2 \Gamma\left(3 - \gamma, {h\nu_l \over k\T_p}\right)  \\ + \xi_3 \Gamma\left(2 - \gamma, {h\nu_l \over k\T_p} \right) \Bigg]^{-1}. 
\end{multline}
The asymptotic behaviour of the incomplete $\Gamma$ function is the following 
\begin{equation}
    \Gamma(s, z\rightarrow \infty) \sim z^{s-1} e^{-z},
\end{equation}
and so 
\begin{equation}
    \eta_X \simeq {2\pi^4 \over 15 \xi_1} \left({h\nu_l \over k\T_p} \right)^{\gamma-3} \exp\left({h\nu_l \over k\T_p} \right) \simeq 5.93 \,  \left({h\nu_l \over k\T_p} \right)^{\gamma-3} \exp\left({h\nu_l \over k\T_p} \right), 
\end{equation}
for low disc temperatures $k\T_p \ll h\nu_l$. As a TDE disc cools with time, the bolometric correction of its light curves will grow exponentially. 

{This equation demonstrates that because the X-ray portion of a typical TDE disc spectrum is in the Wien-tail, the integrated X-ray energy may only correspond to an {\it exponentially suppressed} fraction of the total disc energy. Rather simply this results from X-ray observations only probing the very highest energy photons emitted from the disc, this number of photons decreases exponentially for all energies above the peak disc scale. 

This suppression fraction is, for temperatures $\sim 100$ eV, of order 10 (depending on $\gamma$), and can be as high as 100 for $k\T_p \sim 50$ eV, both typical TDE disc temperatures. A magnitude $\sim 100$ correction is typically what is required to resolve the ``missing energy problem'' of an observed  X-ray bright TDE. The extreme limit of the effect discussed here corresponds to optically-bright TDEs which have peak disc temperatures $\sim 25$ eV, and which are completely unobservable at X-ray frequencies.  }

\begin{figure}
    \centering
    \includegraphics[width=\linewidth]{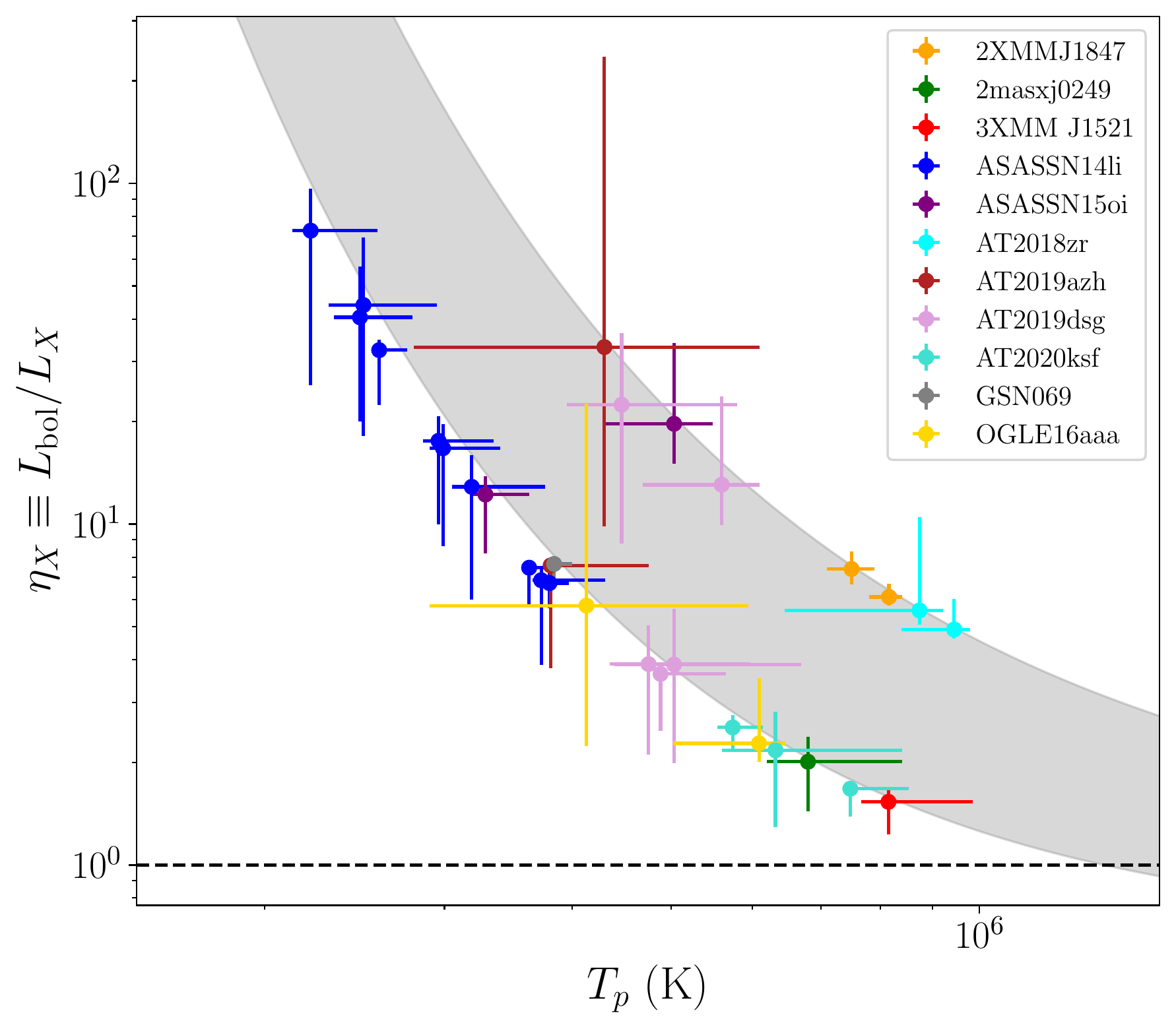}
    \caption{The X-ray ``bolometric correction'' factor, which relates the observed disc X-ray luminosity to the discs bolometric luminosity. The bolometric correction of the TDEs examined in this study ranges from $\sim2$ up to $\sim 100$, meaning that the radiated energy inferred from the X-ray light curve will dramatically underestimate the total radiated energy.  The grey shaded region represents the analytical approximation of equation \ref{etax}, for $1/2 \leq \gamma \leq 3/2$. Note that the bolometric correction of ASASSN-14li grows exponentially at late times.   }
    \label{fig:etaX}
\end{figure}

In Figure \ref{fig:etaX}, we plot the (numerically calculated) bolometric correction of the TDEs in this study against the peak temperature of the disc inferred from each X-ray spectra. The bolometric correction of the TDEs examined in this study ranges from $\sim2$ up to $\sim 100$, meaning that the radiated energy inferred from a TDEs X-ray light curve can dramatically underestimate its total radiated energy. Compounding this effect, the bolometric correction of a TDE disc grows exponentially as the accretion disc cools.   This can be seen in the growth of the bolometric corrections of ASASSN-14li  and AT2019dsg, the best temporally sampled TDEs in our study.   Both sources have bolometric corrections which grow by roughly an order of magnitude over the course of many observations.   

\begin{figure}
    \centering
    \includegraphics[width=\linewidth]{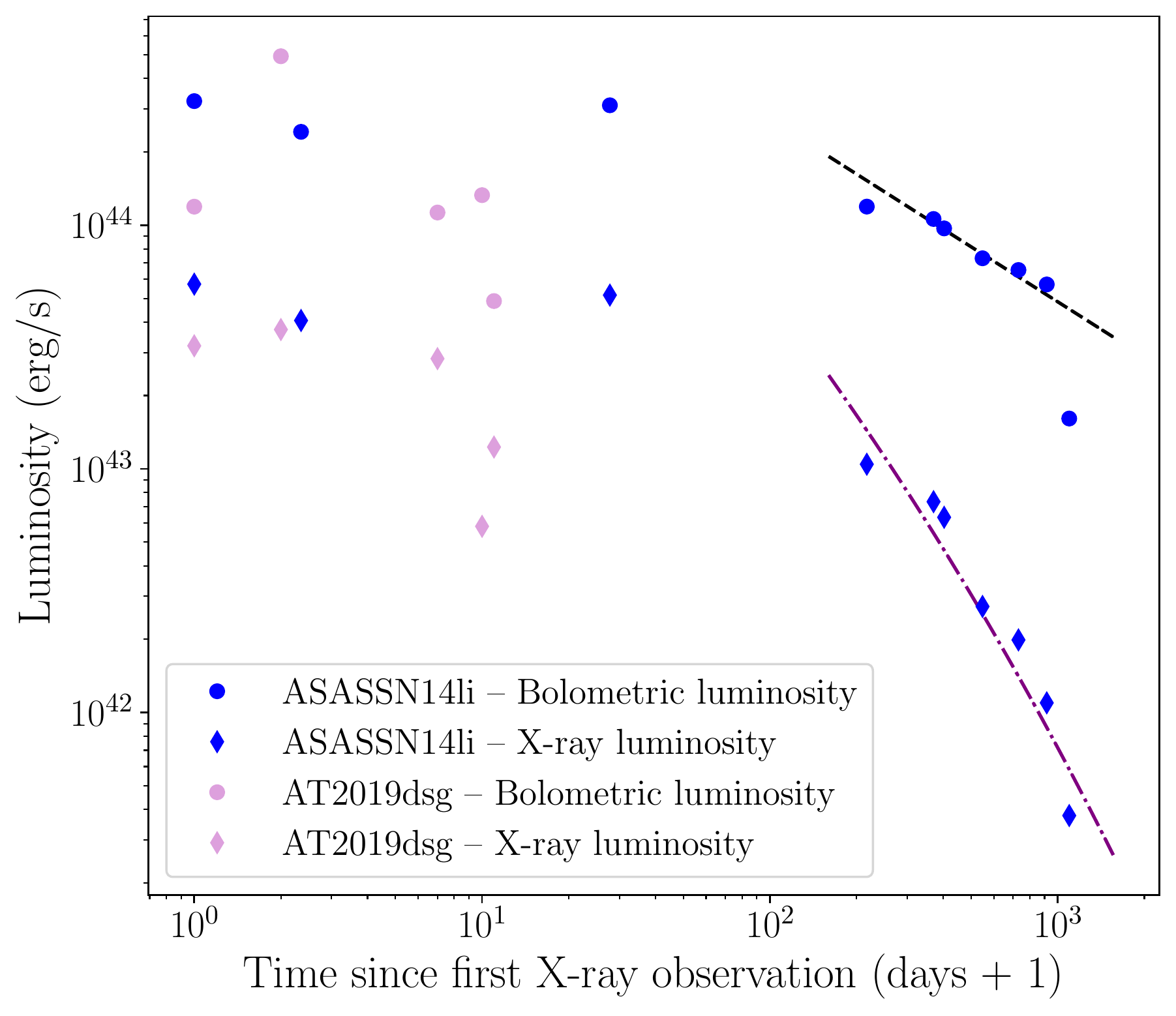}
    \caption{The evolving X-ray (diamonds) and bolometric (circles) luminosity of the two well-sampled TDEs in our sample: AT2019dsg and ASASSN-14li. It is clear to see that the bolometric luminosity of a TDE is always significantly larger than its observed X-ray luminosity. At late times this bolometric correction grows exponentially (as is particularly apparent for ASASSN-14li). The black dashed and purple dot-dashed late time evolution profiles are the theoretical predictions of Mummery \& Balbus (2020), see text. }
    \label{fig:example_lcs}
\end{figure}

The total radiated energy of AT2019dsg's bolometric light curve was a factor 10 higher than it's X-ray light curve, reaching a value $E_{{\rm rad}, {\rm 19dsg}} \simeq 2 \times 10^{50}$ erg $\sim 10 E_{{\rm rad}, {\rm X-ray}}$. However, our observations only span a temporal baseline of roughly 10 days, and a total radiated energy budget is hard to extrapolate from this data set.  For ASASSN-14li however, we have observations spanning more than 1200 days, and a robust estimate of the total radiated energy can be determined.  We find 
\begin{equation}
    E_{{\rm rad}, {\rm 14li}} \simeq 1.17 \times 10^{52} \,\, {\rm erg},
\end{equation}
which corresponds to an accreted mass 
\begin{equation}
    M_{\rm acc} \simeq 0.11 \left({ 0.057 \over \eta}\right) \, M_\odot   , 
\end{equation}
where $\eta$ is the mass to light efficiency of the accretion process ($\eta = 0.057$ is the value appropriate for thin disc accretion onto a Schwarzschild black hole). An accreted mass value of this magnitude is exactly as would be expected from the tidal disruption of a star of mass $M_\star \sim 0.2 M_\odot$, and there is therefore no missing energy. This high value for ASASSN-14li's accreted mass is comparable to those values found from significantly more complex models of ASASSN-14li's evolving light curves, namely the results of  Mummery \& Balbus (2020) and Wen {\it et al}. (2020).

\begin{table*}
    \caption{The maximum early time UV/optical black body luminosity of the six TDEs with excellent early time optical and UV data, compared to the peak of the disc luminosity derived in this paper (i.e., the maximum bolometric luminosity of all the epochs studied here). For sources where the peak disc luminosity was significantly later than the optical/UV peak (ASASSN-15oi, OGLE16aaa and AT2019azh), we also display the disc luminosity at a time closest to optical peak. The column $\Delta t$ denotes the time offset between peak optical/UV and disc luminosity measurements (positive indicating the optical/UV proceeds the disc measurement). Four of the six TDEs have early time optical/UV luminosities in excess of their peak disc luminosities, suggesting that the early time UV/optical emission cannot be solely powered by the reprocessing of disc emission.  }
    \centering
    \begin{tabular}{cccccc}
       Source & $L_{{\rm BB,  max}}$ & $L_{\rm disc, max}$ & $L_{{\rm BB,  max}}\big/L_{\rm disc, max}$ & $\Delta t $& Reference \\
        & (erg/s) & (erg/s) &  & (days) & \\
        \hline
        ASASSN-14li & $1.0 \times 10^{44}$ & $3.1 \times 10^{44}$ & 0.32 & 41 & Holoien et al. 2016a \\ \hline 
        ASASSN-15oi & $1.3 \times 10^{44}$ & $4.8 \times 10^{43}$ &2.7 & 234 & Holoien et al. 2016b \\
        "" & "" & $1.45 \times 10^{43}$ & 9.0 & 76 & "" \\ \hline
        OGLE16aaa & $2.1 \times 10^{44}$ & $2.1 \times 10^{44}$ & 1.0 & 315 & van Velzen et al. 2021 \\
        "" & "" & $5.6 \times 10^{43}$ & 3.75 & 141  & "" \\ \hline
        AT2018zr & $5.6 \times 10^{43} $ & $3.7 \times 10^{42}$ &15.1 & 40 & van Velzen et al. 2021 \\ \hline 
        AT2019azh & $2.8 \times 10^{44}$ & $2.7 \times 10^{43}$ & 10.4 & 200 & van Velzen et al. 2021 \\
        "" & "" & $9.0\times 10^{42}$ & 31.1 & 30 & "" \\ \hline 
        AT2019dsg & $2.9 \times 10^{44}$ & $4.9 \times 10^{44}$ & 0.59 & 18 & van Velzen et al. 2021\\ \hline 
         
    \end{tabular}
    \label{tab:uv}
\end{table*}

The large TDE bolometric correction can be seen most explicitly in Figure \ref{fig:example_lcs}. Here we plot the bolometric (circular) and X-ray (diamond) light curves of ASASSN-14li and AT2019dsg. The light curves of ASASSN-14li are the most illuminating, as they span the longest temporal baseline. At large times ($t-t_0 > 100$ days) the bolometric luminosity of ASASSN-14li falls off by a factor $\sim 5$, while its observed X-ray luminosity falls off by a factor $\sim 150$. As a test that the evolution of both light curves is being driven by the cooling of an accretion disc of fixed area, we plot the two asymptotic theoretical light curve models of Mummery \& Balbus (2020), namely 
\begin{equation}
    L_{\rm bol} \propto t^{-n},
\end{equation}
and
\begin{equation}
    L_X \propto t^{-n/2} \exp\left(- A \left({t + t_0\over t_0}\right)^{n/4} \right) ,
\end{equation}
where 
\begin{equation}
    A_{14{\rm li}} \equiv {0.3 {\rm keV} \over k T_{p, 14{\rm li}}} \simeq 5.75 .
\end{equation}
These expressions are derived under the assumption that the only time dependence inherent in the TDE system is the peak disc temperature cooling according to
\begin{equation}
    T_p \propto t^{-n/4}, \quad n \simeq 0.75 .
\end{equation}
We see in Fig. \ref{fig:example_lcs} that both theoretical profiles produce an excellent description of the evolution.

Clearly, the low values of the observed radiated energy inferred from X-ray observations do not preclude large masses from being accreted onto the central black hole. Instead, the bolometric disc luminosity remains exponentially larger than the X-ray luminosity as the disc cools, and the majority of the energy content is released into far UV frequencies (where the disc spectrum peaks), which are not observed.

\subsection{The early time optical/UV emission}
\begin{figure}
    \centering
    \includegraphics[width=\linewidth]{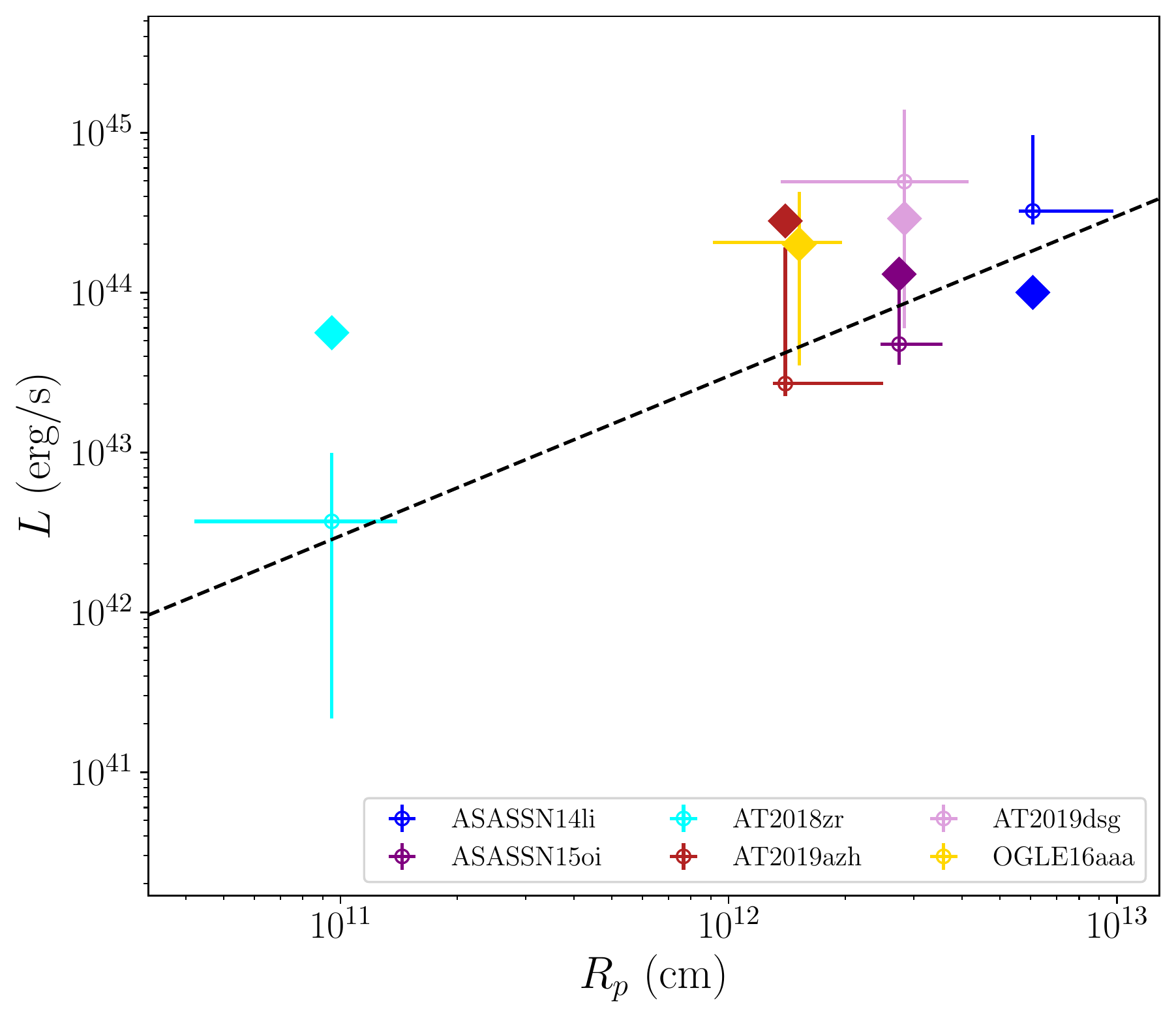}
    \caption{The maximum disc (circular) and UV/optical black body (diamonds) luminosities of the six TDEs in this sample plotted against radial size inferred from the X-ray spectrum (error bars may be smaller than the marker sizes). While the disc luminosity appears to  correlate with the X-ray emission radius (and therefore black hole mass), the UV/optical luminosity does not. In addition, the UV/optical blackbody luminosity exceeds the disc luminosity for four of the six sources. This suggests that the optical/UV luminosity is not produced by reprocessing of the disc luminosity.  }
    \label{fig:uvdisc}
\end{figure}

TDEs are often observed to be extremely bright at optical and UV frequencies at early times.  This early time UV/optical emission typically decays away over $\sim$ years (e.g., Holoien et al. 2016a, 2016b), and is not produced by direct emission from an accretion disc.  At later times (typically a few hundred days post initial disruption),  this early emission has decayed away and the remaining observed optical/UV emission is dominated by {direct} emission from the disc. This direct disc UV emission is characterised  by a much slower, near time-independent, evolution (van Velzen et al. 2019, Mummery \& Balbus 2020, Mummery 2021b).  

The physical origin of this early time emission is contested, with some models assuming that the emission is produced by the reprocessing of accretion disc luminosity by an outflowing photosphere (e.g., Metzger \& Stone 2016, Dai et al. 2018, Nicholl et al. 2020), or by reprocessing from a cooling envelope (Metzger 2022). Other models posit that this emission is produced by shocking debris streams in the disc formation process (e.g., Shiokawa et al. 2015, Piran et al. 2015, Bonnerot \& Stone 2021, Bonnerot et al. 2021).  

One way in which the physical origin of the early time emission can be probed using the results of this study is by comparing the peak of the luminosity of this early time UV/optical component with the peak of the accretion disc luminosity inferred from X-ray spectral measurements. A fundamental requirement for any accretion powered scenario is an ionizing (accretion disc)  luminosity that is equal to or larger than the reprocessed (UV/optical) component. This is something that is directly testable with our new results, as we now have estimates of the intrinsic accretion disc luminosity of each TDE source. In Table \ref{tab:uv} we collate the peak bolometric disc luminosity and the peak `black body' luminosity found from the early time optical/UV emission of the six sources with excellent early time optical/UV coverage from the literature.  This optical/UV luminosity value results from the fitting of a single temperature blackbody to the early time optical emission (which usually provides an acceptable fit to the observations), which is then integrated over a broad range of frequencies (typically corresponding to wavelengths of 0.03-3 microns) to produce a luminosity.   For sources where the peak disc luminosity was significantly later than the optical/UV peak (ASASSN-15oi, OGLE16aaa and AT2019azh), we also display the disc luminosity at a time closest to optical peak. The column $\Delta t$ denotes the time offset between peak optical/UV and disc luminosity measurements. 

From the comparison in Table \ref{tab:uv}, we see that four of the six TDEs have early time optical/UV `blackbody' luminosities  in excess of their peak disc luminosities, suggesting that the early time UV/optical emission cannot be solely powered by the reprocessing of disc emission, and that some additional source of energy input is required. The analysis in Table \ref{tab:uv} is likely to be somewhat conservative, as clearly not all of the accretion disc luminosity can power early optical/UV emission: these sources are detected at X-ray energies.  We reiterate that the density of neutral intervening material found for the sources in our sample is low $n_H \ll 10^{22}$ cm${}^{-2}$, and so it is unlikely that this effect can be explained by our observations ``missing'' some of the intrinsic disc luminosity.

There are other, more circumstantial, lines of reasoning that suggests that the early optical/UV luminosity is not sourced by reprocessed disc emission. The evolutionary properties of the disc and early emission are often substantially different. ASASSN-14li has, for example, a bolometric disc luminosity that varies by a factor $\sim 5$ over $\sim 1000$ days (Fig. \ref{fig:example_lcs}), while it's optical/UV luminosity displays  a much more pronounced decay (Holoien et al. 2016a). 
In addition, the early time optical/UV luminosity of this sample does not correlate with the radius inferred from the X-ray spectral fit (i.e., the sources black hole mass; although see Hammerstein et al. (2022) for evidence of a correlation between peak blackbody luminosity and host galaxy mass), while the peak bolometric disc luminosity does (Figs. \ref{fig:bolmsig}, \ref{fig:bolrad}, \ref{fig:uvdisc}). It seems unlikely that the observed reprocessed accretion luminosity would have fundamentally different scaling properties than the accretion luminosity it is sourced from.

It seems likely therefore that the shocking of debris streams will be important in powering at least some of the early time optical/UV emission, for at least some TDEs.  

\section{Conclusions}\label{conc}
In this paper we have analysed a uniform sample of TDEs observed at X-ray energies, with a new X-ray spectral disc model with physically interpretable parameters.  Of our initial sample of 19 X-ray bright TDEs, 11 inhabit a region of parameter space where this model is valid and we had sufficient data for a detailed analysis of their properties. 

Our key results are the following:\\ (1) The bolometric luminosity of these TDE accretion discs is limited by the Eddington luminosity of their host black holes. A strong linear correlation between peak bolometric luminosity and $M_{\rm BH}-\sigma$ mass is found, indicating that thermal X-ray bright TDE discs form at near universal Eddington luminosity ratios. Quantitatively this  Eddington luminosity ratio was found to be $f_{\rm edd} \sim 0.4^{+0.5}_{-0.2}$. 

(2) This correlation can not be explained by a systematic increase in the neutral absorbing column of low black hole mass TDEs (as may be expected if lower mass BHs have highly super-Eddington accretion rates, driving strong accretion disc winds), as the column depth of the sources in our sample do not correlate with black hole mass. We find low levels of neutral intervening material $n_H \ll 10^{22}$ cm${}^{-2}$ for all TDEs of our sample. 

(3) This correlation is robust, and does not depend on the use of the $M_{\rm BH}-\sigma$ relationship in computing the TDEs black hole masses. The radii inferred from X-ray spectra of our TDE sample lie in the parameter space expected for the ISCO radii of black hole masses corresponding with the observed galactic velocity dispersion (Figs. \ref{fig:fig2}, \ref{fig:fig1}), and can be used to infer an independent black hole mass measurement for each TDE.  Using the mass measurements inferred from the TDE's X-ray radii we again find a linear correlation between bolometric disc luminosity and black hole mass, and disc Eddington luminosity ratios of $\sim 10$\% (Fig. \ref{fig:edratRP},section 5.3).

(4) We demonstrate how the small inferred radiated energies from TDE X-ray light curves can be understood by the large bolometric corrections inherent to their spectra. As TDE discs cool, the bolometric luminosity of their discs remains exponentially higher than their observed X-ray energies, and the X-ray-to-bolometric conversion factor can reach $\sim 100$. The bolometric  radiated energy of ASASSN-14li, the best observed source in our sample, is $\sim 1 \times 10^{52}$ erg ($M_{\rm acc} \sim 0.1 M_\odot$), meaning that it has no ``missing'' energy.  

(5) We demonstrate that the early time optical and UV luminosity of many of the sources in our sample exceed their bolometric disc luminosities by a significant margin.  This  suggests that additional energy sources must be present at early times, and that the early optical emission seen from some TDE sources cannot be explained by the reprocessing of accretion disc emission.\\  

It is important to remember that our conclusions are a result of an analysis of TDEs observed to have bright, thermal, X-ray spectra, and this must be kept in mind when generalising these conclusions to the entire TDE population. 

This work does however provide strong evidence that thermal X-ray TDEs behave as ``scaled up'' analogues of Galactic X-ray binaries in the soft state, and that at least some black hole accretion states are black hole mass independent. 

The large TDE samples expected to be uncovered with all sky X-ray surveys such as eROSITA provide a unique opportunity to probe the occupation fraction and properties of black holes at the low mass end of the galaxy/SMBH function. In the near-future, the bottleneck in constraining SMBH demographics with TDEs will shift from the sample size to a lack of adequate follow-up observations. Here we have shown that with only a few 100 soft X-ray photons (readily available in most X-ray observations), it is possible to estimate the SMBH mass following a TDE. This will provide a complementary avenue to TDE optical light curve models, many of which are not self-consistent and/or incorporate phenomenological physics to make up for our lack of understanding of the UV/optical emission mechanism. 

\section*{Data availability statement}
The data used in this manuscript will be made available on Zenodo post publication. The Zenodo doi is 10.5281/zenodo.7533374.  An XSPEC-ready  implementation of the disc fitting function is available at this url: \href{https:www.github.com/andymummeryastro/TDEdiscXraySpectrum}{github.com/andymummeryastro/TDEdiscXraySpectrum}.


\section*{Acknowledgements} 
This work was supported by a Leverhulme Trust International Professorship grant [number LIP-202-014]. For the purpose of Open Access, AM has applied a CC BY public copyright licence to any Author Accepted Manuscript version arising from this submission. 
TW warmly thanks the Space Telescope Science Institute for its hospitality during the completion of this work.
Based on observations obtained with XMM-Newton, an ESA science mission with instruments and contributions directly funded by ESA Member States and NASA.
This paper includes data gathered with the 6.5 meter Magellan Telescopes located at Las Campanas Observatory, Chile (PI: Pasham).
We acknowledge the use of public data from the Swift data archive.

\appendix{}
\section{Observation ID'{s}  }
In Table \ref{table:1} we collate the observational ID's, instrument used and observation dates of all of the X-ray spectra modelled in this work. This includes the sources with either too low photon counts to be analysed (e.g., AT2019qiz), or disc spectra with prominent power-law components (e.g., XMMSL2 J1446), which were not included in the population analysis in this paper. 

\begin{table*}
\caption{Overview of the sample and observations. }
\label{table:1}      
\centering          
\begin{tabular}{ccccccc}     
Source & Distance & Instrument & Observation ID & MJD & Photon Counts  \\
 & (Mpc) & & & (days) & &  \\
\hline\hline
ASASSN--14li & 90 & XMM/RGS & 694651201 & 56997.2 & 9200  &  \\
&&XMM/RGS &	722480201&56998.5& 36150  &  \\
&&XMM/RGS	& 694651401&57024& 9600  &  \\
&&XMM/RGS	& 694651501&57213.3 & 2086  & \\
&&XMM/RGS	& 770980101&57366.5 & 5700  & \\
&&XMM/PN&	770980501 & 57399.2& 22400 & \\
&&XMM/PN&	770980601 & 57544.1& 18400  &  \\
&&XMM/PN&	770980701 & 57726.6& 9800  & \\
&&XMM/PN&	770980801 & 57912.1& 8200  & \\
&&XMM/PN&	770980901 & 58092.5& 7600  & \\
\hline
ASASSN--15oi &216 & XMM/PN&	722160501 & 57324 & 600 &   \\
&& XMM/PN & 722160701 & 57482 & 4050 &   \\\hline
OGLE16aaa & 800 & XMM/PN & 790181801 & 57548 &  260 \\
&& XMM/PN & 793183201 & 57722 & 4500 \\\hline 
AT2018zr &322 & XMM/PN&	822040301&58220 & 300 &   \\
&& XMM/PN&	822040501&58242& 185 &   \\\hline
3XMMJ1500 &692 & XMM/PN &	554680201& 54873 & 1700 &   \\
&&XMM/PN	&804370401& 57974 & 160 &  
\\\hline
AT2019dsg & 224 & NICER & 2200680101 & 58624 & 13000 \\
&& NICER & 2200680102 & 58625 & 3800 \\
&& NICER& 2200680103 & 58630 & 1300  \\
&& NICER & 2200680105 & 58633 & 750 \\ 
&& NICER & 2200680106 & 58634 & 660 \\ \hline
AT2018fyk & 264 & Swift && 58383 - 58446  & 1400 \\
&& XMM/PN & 831790201 & 58461.75 & 20200 \\
&& XMM/PN & 853980201 & 58782.21 & 33000 \\ \hline 
XMMSL1 J0740 & 73 & XMM/PN & 740340401 & 56777 & 7100 \\ 
&& XMM/PN & 740340601 & 57033 & 3600 \\ 
&& Swift & 33229001 - 33229005 & 56758 - 56796 & 810 \\ \hline 
3XMM J1521 & 866 & XMM/PN & 109930101 & 51778 & 3050 \\ \hline 
GSN069 & 79 & XMM/PN & 740960101 & 56996 & 55000 \\ \hline
SDSSJ1201 & 700 & XMM/PN & 555060301 & 55369 & 2600 \\
&& XMM/PN & 555060401 & 55523 & 960 \\ \hline 
XMMSL2 J1446 & 127 & XMM/PN & 763640201 & 57609 & 600 \\ \hline 
2XMMJ1847 & 156 & XMM/PN & 405380501 & 53985 & 18000 \\
&& XMM/PN & 405550401 & 54206 & 2000 \\ \hline 
AT2018hyz & 204 & Swift & &58432 - 59000& 50 \\ \hline 
RBS1032 & 114 & XMM/PN & 604020101 &55156& 120 \\ \hline 
AT2019azh & 96 & Swift & & 58553 - 58634 & 240 \\
&& Swift && 58767 - 58977 & 2500 \\ \hline 
2MASXJ0249 & 83 & XMM/PN & 411980401 & 53930& 1780 \\ \hline 
AT2019qiz & 66 & Swift & 12012043 &59623 & 65 \\ \hline 
AT2020ksf & 426 & NICER &3639010101 &  59187 - 59189 & 14450 \\
&& NICER & 3201930101 & 59191 - 19195 & 11230 \\ 
&& Swift  & & 59179 - 59205 & 440 \\\hline 
\end{tabular}
\end{table*}

\section{Example X-ray spectra}\label{secB}
\begin{figure*}
    \centering
    \includegraphics[width=0.45\textwidth]{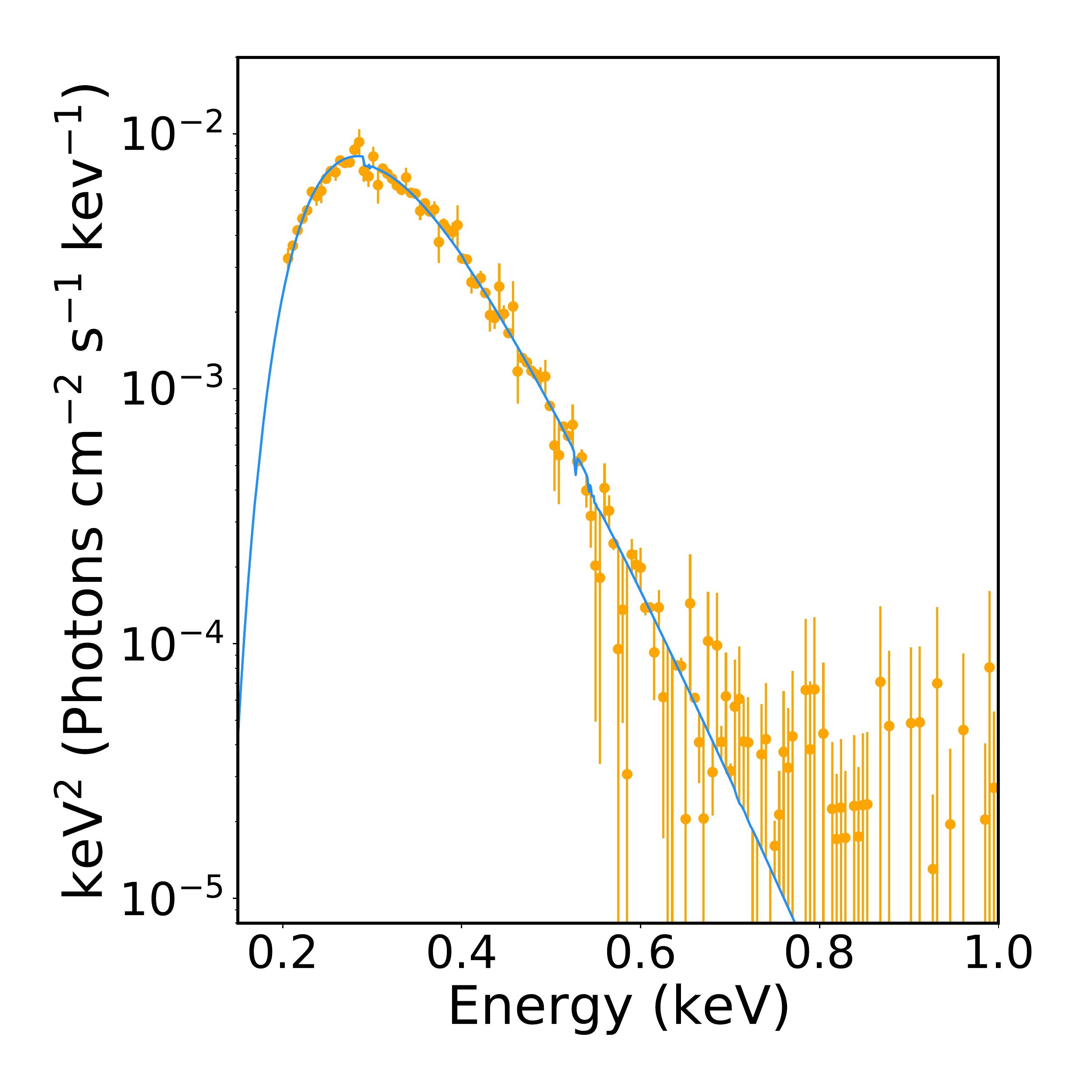}
    \includegraphics[width=0.45\textwidth]{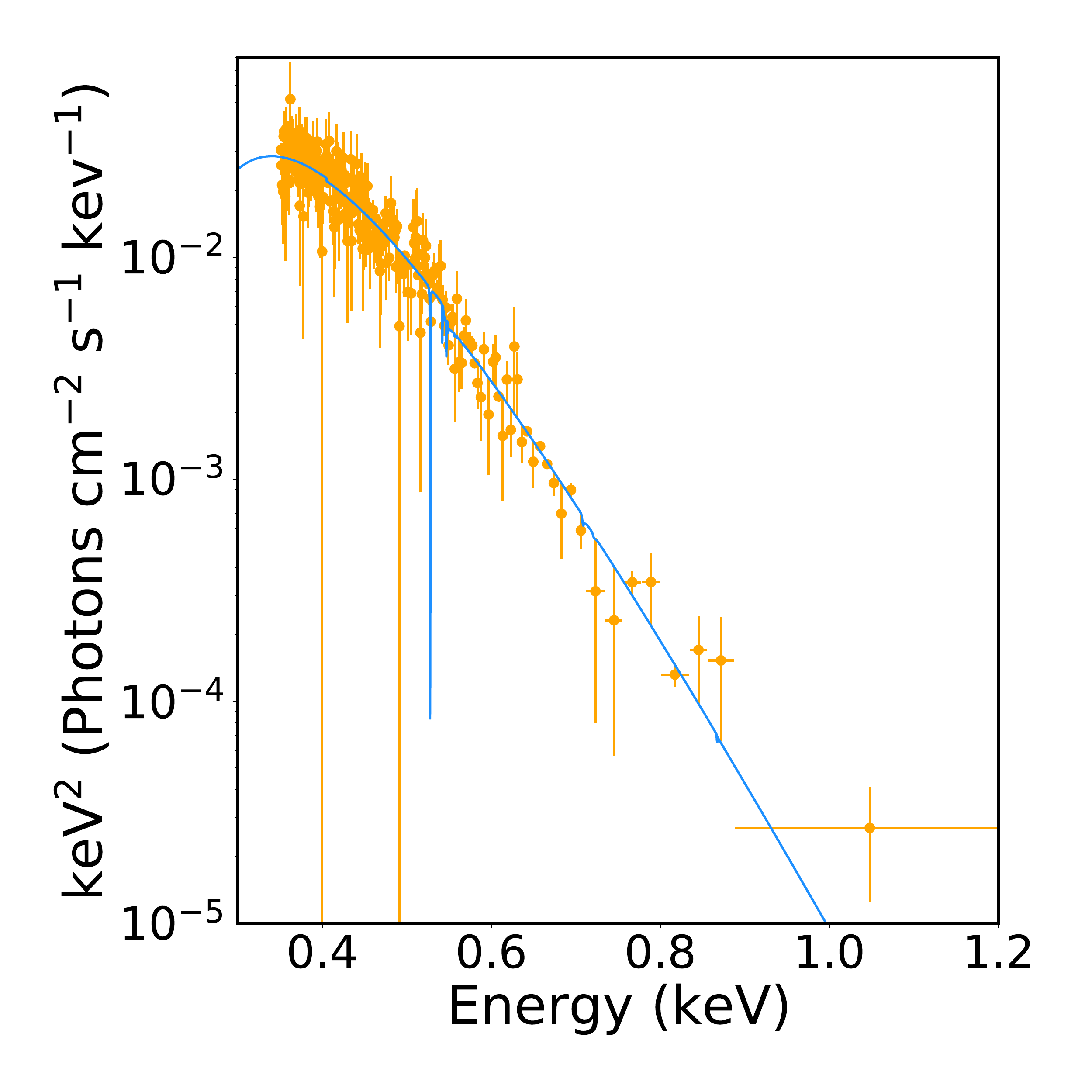}
    \includegraphics[width=0.45\textwidth]{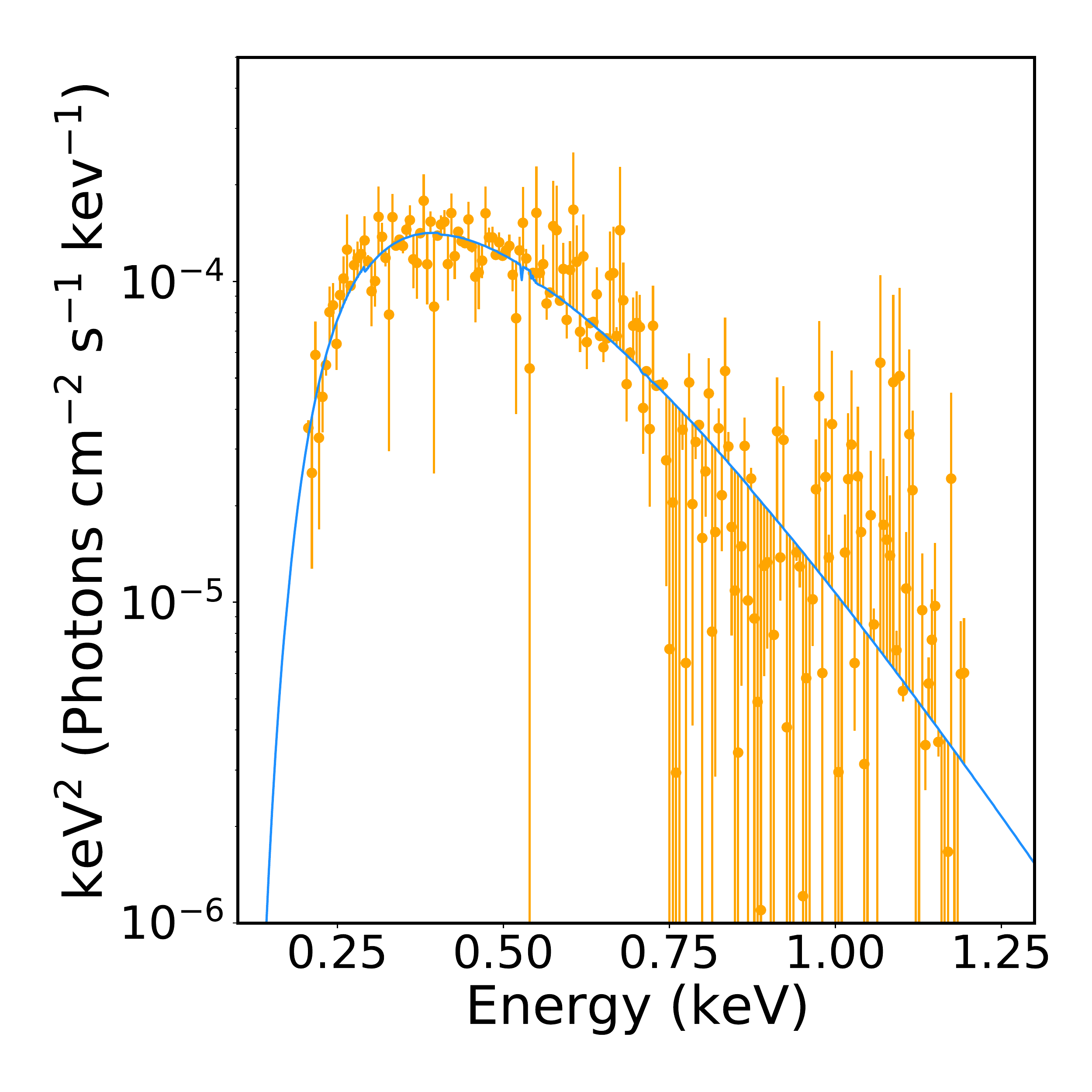}
    \includegraphics[width=0.45\textwidth]{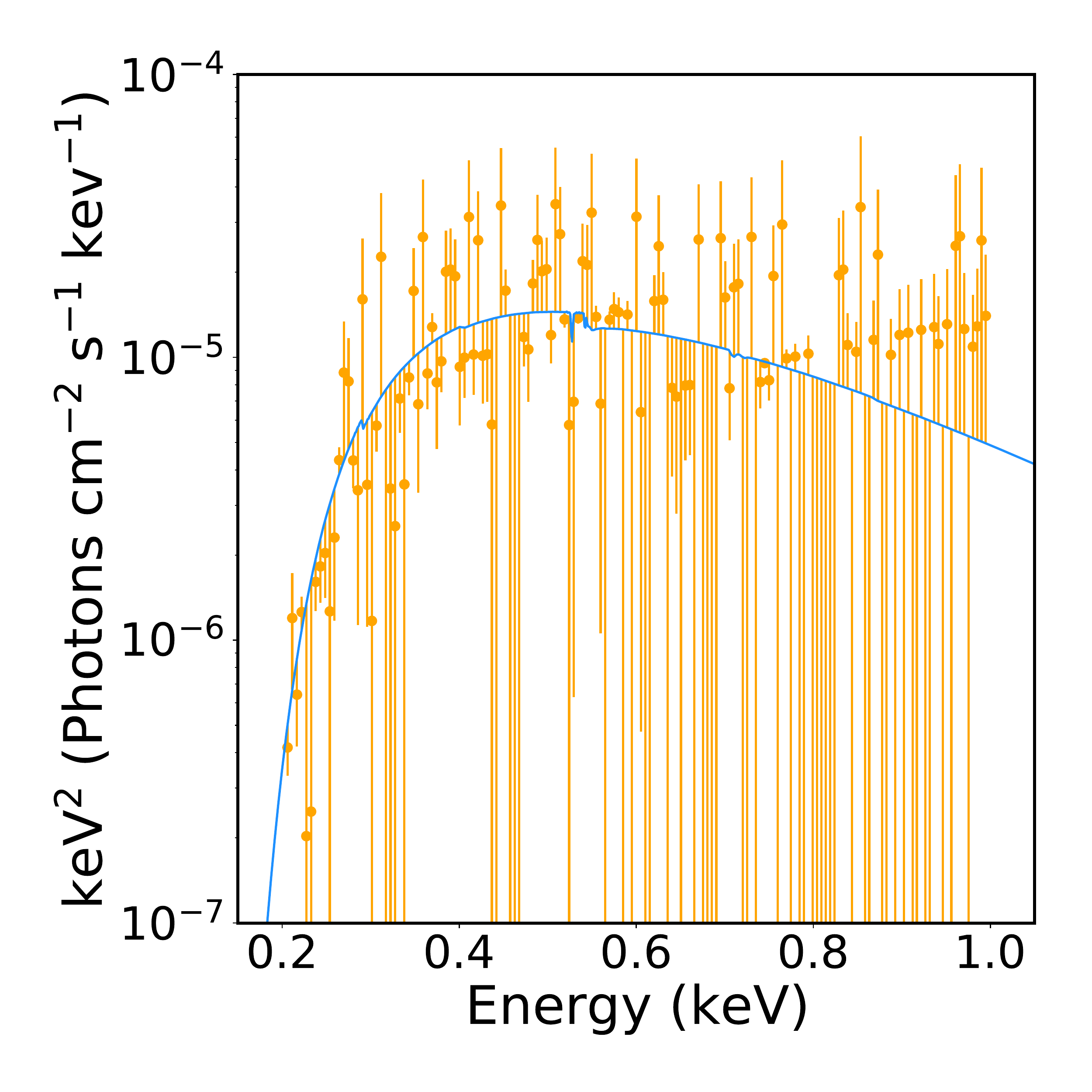}
    \caption{Four examples of X-ray spectra (orange) and their best-fit model ({\tt tdediscspec $\times$ TBabs}, blue). These examples illustrate different data quality levels, ranging from excellent (top left; 22400 photon PN spectrum of ASASSN--14li at MJD 57399) to good (top right; 9200 photon RGS spectrum of ASASSN--14li at MJD 56997) to moderate (bottom left; 3050 photon PN spectrum of 3XMMJ1521 at MJD 51778) to poor (bottom right; 300 photon PN spectrum of AT2018zr at MJD 58220). These spectra also illustrate the excellent sensitivity to the curvature of spectrum in the PN instrument at low energies.}
    \label{fig:examplespectra}
\end{figure*}

In this Appendix we display some example X-ray spectra of the sources studied in this work (Figure \ref{fig:examplespectra}). 

\section{Bolometric luminosity inferred from the X-ray spectra: testing model assumptions  }
\begin{figure}
    \centering
    \includegraphics[width=\linewidth]{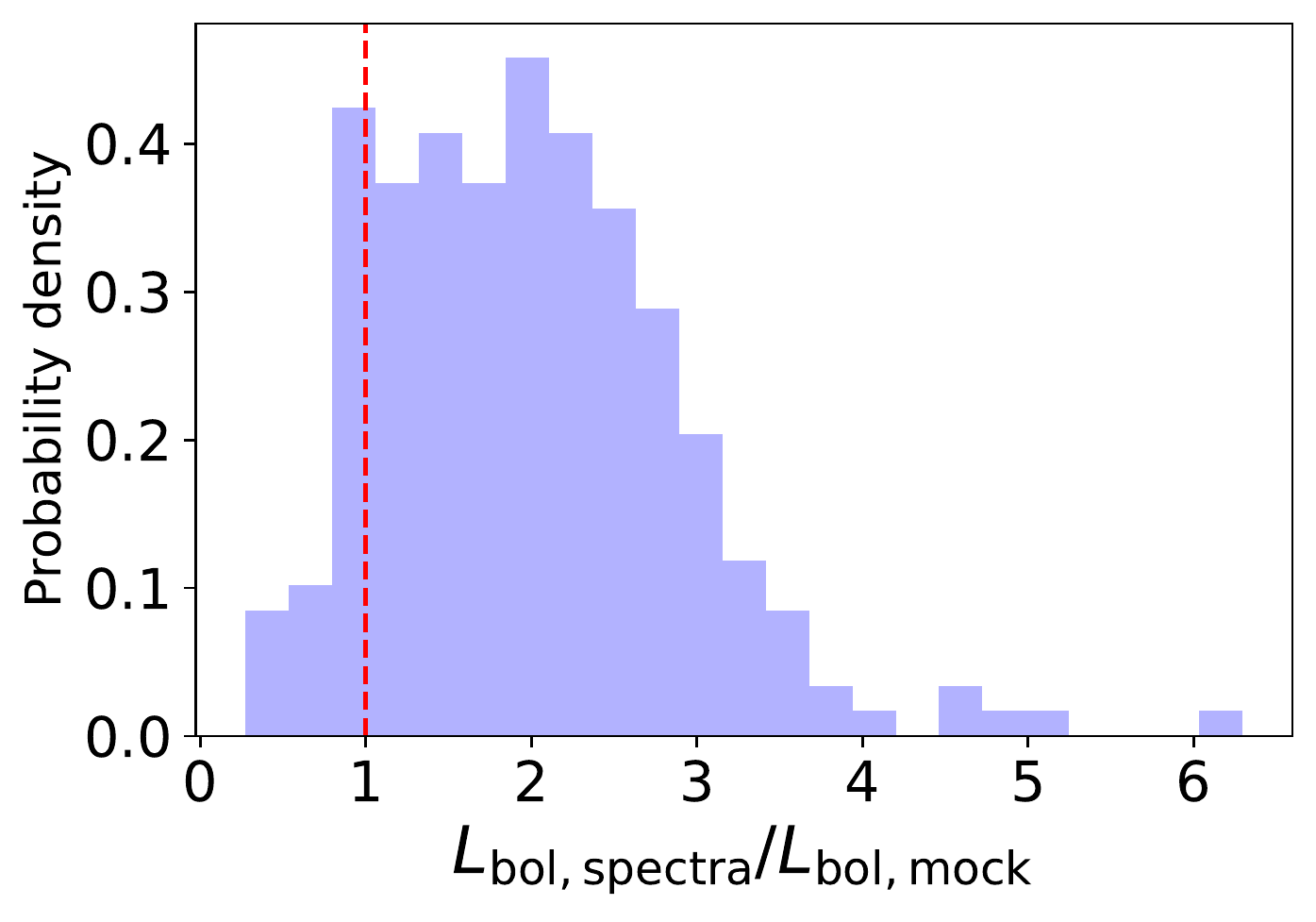}
    \caption{The ratio of the bolometric luminosity inferred from the mock disc spectra and the exact value of the numerical disc solutions. This analysis assumes a uniform distribution of both black hole spins and disc-observer inclination angles. The red vertical dashed line corresponds to 1, i.e., a perfect inference. We see that the models in this paper slightly overestimate the luminosity on average, but by a factor typically much less than the error range introduced by the uncertainty in the fitted parameters.  }
    \label{C1}
\end{figure}

{In this section we test the model assumptions used in the derivation of equation (\ref{LBOL}). We do this following the procedure of section 5.3: we numerically simulate and fit a number of 0.3--2 keV X-ray spectra with known black hole masses, spins and inclinations with the model of equation \ref{MB}. The mock X-ray spectra are generated by solving the relativistic thin disc equations, and ray-tracing the resulting temperature profiles (see e.g., Mummery \& Balbus 2020 for a description of the algorithms used to solve both the disc evolution and photon orbit equations). The mock spectra were then fit with equation \ref{MB}, and the parameters $R_p$ and $T_p$ are used to compute $L_{\rm bol}$ from equation \ref{LBOL}, which was then compared to the exact value of the bolometric disc luminosity of the numerical solutions. The ratio of these two quantities, calculated for a uniform distribution of both black hole spin and disc observer inclination angles is displayed in Fig. \ref{C1}. We see that the models in this paper slightly overestimate the luminosity on average, but by a factor typically much less than the error range introduced by the uncertainty in the fitted parameters.  }

\label{lastpage}

\end{document}